\documentclass[aps,prl,showpacs,amsmath,amssymb,
	preprint,onecolumn]{revtex4}

\usepackage[dvips]{graphicx}
\usepackage{dcolumn}
\usepackage{bm}
\usepackage{epsf}

\begin{document}

\title{Zero Lag Synchronization of Mutually Coupled Lasers in the Presence of
  Delays}

\author{Alexandra S. Landsman$^1$, Leah B. Shaw$^2$, and Ira B. 
Schwartz$^1$}
\affiliation{$^1$Nonlinear Systems
  Dynamics Section, Plasma Physics Division, Naval Research 
Laboratory, Washington, DC 20375.
$^2$Department of Applied Science, College of William and Mary, Williamsburg,
VA 23187.}

\begin{abstract}

We consider a line of three mutually coupled lasers with time delays
and study chaotic synchronization of the outer lasers.  Two different
systems are presented:  optoelectronically coupled semiconductor lasers
and optically coupled fiber lasers.  While the dynamics of the two systems
are very different, robust synchronization of end lasers is obtained in
both cases over a range of parameters.  Here, we present analysis and
numerical simulation to explain some of the observed synchronization
phenomena.  First, we introduce the system of three coupled semiconductor
lasers and discuss the onset of oscillations that occurs via a bifurcation
as the coupling strength increases.  Next, we analyze the synchronization
of the end lasers by examining 
the dynamics transverse to synchronized
state.  We prove that chaotic synchronization of the outer semiconductor
lasers will occur for sufficiently long delays, and we make a comparison
to generalized synchronization in driven dissipative systems.  It is shown
that the stability of synchronous state (as indicated by negative Lyupunov
exponents transverse to the synchronization manifold) depends on the
internal dissipation of the outer lasers.  We next present numerical
simulations for three coupled fiber lasers, highlighting some of the
differences between the semiconductor and fiber laser systems. Due to the
large number of coupled modes in fiber lasers, this is a good system for
investigating spatio-temporal chaos.  Stochastic noise is included in the
fiber laser model, and synchrony of the outer lasers is observed even
at very small coupling strengths.

\end{abstract}


\maketitle

\section{Introduction}
When two or more systems are coupled, their interaction often leads to
correlations in the dynamics.   
If the dynamics of these coupled systems are identical with respect to 
some measure, 
the correlated motion is considered synchronized. Synchronization
has been studied since the time of Christian Huygens, and there now
exist several reviews on the dynamics of synchrony in the literature,
such as Refs.~\cite{BoccalettiKOVZ02,pikovsky,DingDDGIPSY97,Strogatz00}.
It is now clear that synchronization appears in a wide range of applications
from many fields of science, such as physics, engineering, biology
and chemistry, as well as in various fields of social behavior. 

In general, synchronization between two interacting systems may be
quantified by examining and comparing the output time series from
each dynamical system. Typically, a measure of correlation between
 signals may be used to classify the type of synchronization.
Complete synchronization occurs in coupled phase oscillators \cite{Strogatz00}
as well as in coupled chaotic oscillators \cite{FujisakaY83,PecoraC90}.
In this case, amplitudes and phases are identical, and the peak of
the cross correlation between the signals is at zero time shift. A recent
theoretical example 
of complete synchrony in a closed ring of three one dimensional Ikeda
oscillators with delayed diffusive coupling has been seen in \cite{BuricTV07},
and complete synchrony has been shown in two mutually  delay coupled
lasers with self feedback for models of both semiconductor lasers and 
fiber ring lasers \cite{IBSLBSciteulike:1222889}.  

Many dynamical phenomena beyond complete synchronized systems have been
unveiled 
for two coupled systems. If the amplitudes are uncorrelated but the
phases are locked, or entrained, between the two signals, then the
systems are said to be in phase synchrony \cite{PikovskyRK96}. One
other type of synchronization, called \emph{generalized synchronization}, 
deals solely with the unidirectional
coupling between two oscillators of drive and response type 
\cite{RulkovSTA95}. In generalized
synchronization, there exists a functional relationship between the
drive and response, where there exists a function $F$ such that
$\mathbf{X}_{2}(t)=F(\mathbf{X}_{1}(t)),$ where $X_1$ and $X_2$ are
the time series for the driver and the response systems, respectively.
In another setting, this may
also be thought of as a generalized entrainment in dynamics, whereby
one system is entrained functionally to another. Many examples of
entrained systems occur in singularly perturbed problems, and specifically
in systems with multiple time scales where dimension reduction forces
a functional relationship to occur between dependent and independent
coordinates\cite{SchwartzMBL04}. Generalized synchronization also
plays a crucial role in mutually coupled systems with long delays
\cite{LandsmanS07}, where the coupling term can be viewed as a
driving signal over the interval of the round-trip time.  This idea
will be further elaborated in the present chapter in connection to
chaotic synchronization of semiconductor lasers with long coupling
delays. 


One area used to explore interesting synchronization phenomena in delay
coupled systems is that of nonlinear optics. Coupled lasers, both
semiconductor lasers as well as spatio-temporal fiber lasers, have
been used to study delay coupled dynamics experimentally as well as
theoretically. In  delay coupled systems, a time lag between
the oscillators is typically observed in the cross correlation, with
a leading time series followed by a lagging one. Such lagged systems
are defined to exhibit \emph{achronal synchronization.} Existence
of achronal synchronization in a mutually delay-coupled semiconductor
laser system was shown experimentally \cite{HeilFEMM01}, and
studied theoretically \cite{WhiteMM02} in a single-mode semiconductor 
laser model.
Anticipatory synchronization occurs when a response in a driven system's
state is not replicated simultaneously but instead anticipates the dynamics
of the driver \cite{Voss00,Voss01}.  An example of anticipatory
synchronization in the presence of delays
can be found in coupled semiconductor 
lasers \cite{Masoller01,SivaprakasamSSS01}, and has also been observed 
in the presence of stochastic effects in models of excitable media
\cite{CiszakCMMT03}.  Interestingly, anticipated synchronization was also
observed in unidirectionally coupled systems with no delays \cite{Voss00} and
was recently studied numerically and experimentally in coupled Rossler
oscillators \cite{Pisarchik06,Pisarchik07}.
The zero lag state, corresponding to complete synchronization, 
is generally unstable in the delay coupled systems 
where achronal or anticipatory behavior is observed.  Moreover, when
achronal synchronization occurs, the situation may be further complicated
by switching between leader and follower \cite{MuletMHF04,ShawSRR06}. 
We are particularly interested in systems where the zero lag state, also known
as isochronal state, is stable.  
The stability of this isochronal solution is often due
to a coupling geometry that leads to complete synchronization in
a variety of systems \cite{Landsman07}.  

The present chapter focuses on 
optoelectronically coupled semiconductor lasers and fiber ring lasers as
important examples of systems exhibiting isochronal synchronization when
coupled in a certain way and in the presence of delays.  
This isochronal synchronization can be compared to the above discussed
achronal synchronization of mutually coupled lasers, for which the 
solutions are identical, but shifted in time with respect to
each other \cite{Chiang05,Chiang06}.
Previously, there has been some investigation of chaotic synchronization
of the outer lasers mutually coupled in a line \cite{TerryTDVZAR99,Yanchuk06}
in the absence of delays.  Delays, however significantly complicate the
analysis by possibly introducing an infinite number of degrees of freedom
into the system.
In general, semiconductor lasers are considered low dimensional,
since they are modeled by differential equations with no spatial
component due to very short cavity lengths. In contrast, fiber lasers
have light propagating over long distances through  a length of optical 
fiber, forming
a large number (on the order of $10^{3}$) of longitudinal modes.
Thus even a single fiber ring laser  exhibits high dimensional
spatio-temporal dynamics. The chaotic dynamics of fiber ring lasers
have been studied in the past. Experiments exploring  the 
polarization mode dynamics in a single fiber ring laser were set up and 
modeled using a
delay differential system in \cite{WilliamsR96}. Other experiments 
on synchronization
with coupled fiber lasers have been reported in 
\cite{VanwiggerenR99,WangS02,ImaiMI03},
and noise-induced generalized synchronization in fiber ring lasers
has been reported in \cite{DeShazerTKR04}. Modeling the ring laser
yields a system of equations which consists of coupled difference
and differential delay equations. To obtain better agreement with
experiment, it was found that inclusion of spontaneous emission effects
was necessary in the modeling, which resulted in a stochastic
 difference-differential
system of equations \cite{WilliamsGR97}, and this approach was followed
in \cite{ShawSRR06}.

In this chapter we explore complete or isochronal
synchronization in mutually delay coupled
systems.  The coupling architecture is three lasers coupled in a line.  
The layout of the chapter is as follows.  The first two sections 
explore a system of three semiconductor lasers with significant delays.  
We first introduce the 
model and show an instability of the steady state for sufficiently strong 
coupling.  We then study chaotic synchronization that occurs for stronger 
coupling and prove the stability of the synchronized state for sufficiently
long delays.  The following section treats three fiber ring lasers with the 
same coupling architecture and shows synchrony of the end lasers in 
numerical simulation.  We then conclude and discuss possible avenues for 
future work.

\section{Semiconducting Laser Model and Onset of Oscillations}
In this and the following section we will investigate the dynamics of three
semiconductor lasers coupled in a line with delays.  
Previously, the dynamics of two optoelectronically delay coupled lasers have 
been explored,  showing lag synchronization between the two lasers and 
isochronal synchronization if self feedback is
added \cite{VicenteTMML06}.    

\begin{figure}[h]
\hspace*{-1 cm}
{
\epsfxsize=6in
\epsffile{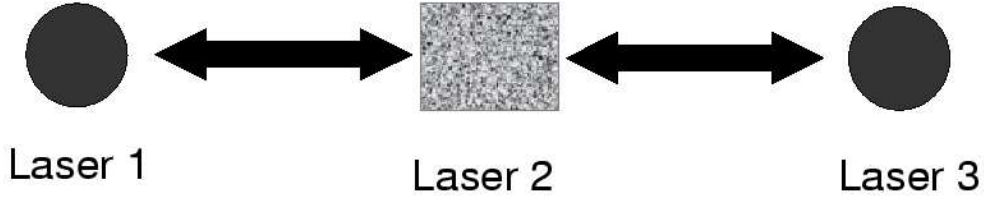}
}
\caption{A schematic showing how three lasers are coupled in a
  line. The outer two lasers (circles) are identical, while the middle
  laser (square) is
  detuned from the rest. }
\label{fig:3lasers}
\end{figure}

A schematic of the set-up is shown in Fig.~\ref{fig:3lasers}.
The coupling is optoelectronic, where
the coupling signal from one laser is transmitted to the other via fiber-optic
cable and an electronic circuit that introduce a time delay.  In the 
absence
of coupling, each laser is tuned so that it emits a stable constant light
output.  In the presence of coupling, there are fluctuations in the light
intensity of the lasers that are converted to an electronic signal which
controls the pump strength of the coupling term \cite{CarrTS06}.    
This type of coupling is called ``incoherent coupling'' since it does
not contain phase information.  Because the phase of the electric field is
not used in coupling one laser to another, the equations of each laser can be
modeled by only two variables; i.e., intensity and inversion.  

The scaled equations of coupled
semiconductor lasers have the following form \cite{KimRACS05,CarrTS06}:
\begin{displaymath}
\frac{dy_1}{dt} = x_1 \left(1+y_1\right)
\end{displaymath}  
\begin{equation}
\frac{dx_1}{dt} = -y_1 - \epsilon x_1\left(a_1 + b_1 y_1\right) + \delta_2
y_2\left(t-\tau\right)
\label{eq:lasers1}
\end{equation}  
\begin{displaymath}
\frac{dy_2}{dt} = \beta x_2 \left(1+y_2\right) 
\end{displaymath}  
\begin{equation}
\frac{dx_2}{dt} = \beta \left[-y_2 
- \epsilon \beta x_2 \left(a_2 + b_2 y_2\right) \right]
+ \delta_1 \left[y_1\left(t-\tau\right) + y_3\left(t-\tau\right)\right]
\label{eq:lasers2}
\end{equation}
\begin{displaymath} 
\frac{dy_3}{dt} = x_3 \left(1+y_3\right) 
\end{displaymath}
\begin{equation}
\frac{dx_3}{dt} = -y_3 - \epsilon x_3\left(a_1 + b_1 y_1\right) + \delta_2
y_2\left(t-\tau\right)
\label{eq:lasers}
\end{equation} 

Variables $y_i$ and $x_i$ denote scaled intensity and inversion
of the $i$th laser, respectively,
$\{a_1,a_2,b_1,b_2\}$ are loss terms, 
and $\epsilon$ is the dissipation.  The subscript on $\delta_2$ signifies that
the coupling is from the middle to the outer lasers.  Similarly,
$\delta_1=\delta_3$ signifies the coupling strength
from the outer to the middle lasers \cite{KimRACS05,CarrTS06}.
Detuning of the middle laser from the outer ones is given by
$\beta$: the ratio of the relaxation frequency of the middle to the outer
lasers.  If we focus only on linear terms in Eqn. (2), then we can
easily recover
the equation for a simple harmonic oscillator: $\ddot y_2 = - \beta^2 y_2$,
which shows the correct dependence of force on the dimensionless frequency squared.

Equations (\ref{eq:lasers1})-(\ref{eq:lasers}) are scaled in such
a way that $y_i > -1$, since the motion slows down asymptotically
as $y_i \rightarrow -1$.  (See \cite{SchwartzE94} for details of the 
derivation
from the original physical model.)  It follows that the $\epsilon$ term
in the above equations is always dissipative, 
provided $a_{(1,2)} > b_{(1,2)}$, 
and leads to a spiraling of the  dynamics towards zero in the absence of mutual
coupling.  In the actual experiment, this zero state would correspond to some
constant steady state output.  The above equations are coupled via
laser intensities, $y_i$, using optoelectronic incoherent 
coupling. 

The delay in the coupling terms is fixed and  given by $\tau$, and
the strength of coupling from the center to the outer identical subsystems by
$\delta_1$, while from the outer to the center by $\delta_2$.  
Variables $\{x_1,y_1\}$ and $\{x_3,y_3\}$ are symmetric with respect to 
interchange of variables.
Due to this internal symmetry of the system, there exists an
identical solution for the outer lasers: 
$x_1=x_3; y_1 = y_3$.  If this solution is stable then the outer 
lasers are synchronized.
In the absence of dissipation ($\epsilon=0$), the uncoupled
system ($\delta_1=\delta_2=0$) is a nonlinear conservative system, 
with behavior similar to a
simple harmonic oscillator for small amplitudes, and becoming more pulse-like
at high amplitudes \cite{SchwartzE94}.  Dissipation, however, leads to 
energy loss, so that in
the absence of coupling, the dynamics would settle into a
steady state: $\{x_i=0, y_i=0\}$. 
Mutual coupling acts like a drive by pumping energy into
the system, similar to a laser with injection. Recent studies of two mutually
coupled semiconductor lasers with delay show explicitly in both theory and 
experiment how the
amplitude of the intensity scales with coupling strength for the case of 
fixed delay \cite{KimRACS05}. For most cases of physical interest, it can 
be assumed that dissipation is small:
($\epsilon \ll 1$).  We assume small dissipation and $a_{(1,2)} > b_{(1,2)}$
throughout the rest of  this chapter.  It can be seen from
Eqns. (\ref{eq:lasers1})-(\ref{eq:lasers}) that at low amplitudes the
relaxation frequency is equal to one, so that the period of a single
oscillation is given by $2 \pi$, in the scaled variables used in the
equations.  In the typical experimental set-up, the relaxation
oscillations are on the order of $2-3$ ns per cycle.  
Since the delay time is set to be about an order of magnitude higher than
the period of oscillation, we use about $\tau=60$ as a typical delay time
in many of the simulations, which corresponds to delays of
about $20-30$ ns in an experimental set-up.

We now explore the
onset of regular oscillations that occurs when the coupling strength between
lasers is above a bifurcation value.  Below this bifurcation value, the
oscillations are damped out to steady state due to dissipation, $\epsilon$, in
the lasers.  It is clear from Eqns. (\ref{eq:lasers1}) - (\ref{eq:lasers}) that
the steady state, $\{x_i=0, y_i=0\}$, is a solution.  To determine the
stability of this solution, we linearize about the steady state, looking 
at time-evolution of small perturbations.  It can be seen from the laser
equations that at small amplitude the linear terms dominate, so that the 
dynamics are close to that of coupled simple harmonic oscillators.  
Since even at small amplitudes, we have a linear
dissipation term, $-\epsilon a_1 x_i$ (see Eqns.  (\ref{eq:lasers1}) -
(\ref{eq:lasers})), the coupling will only induce oscillations if
it contributes enough energy to each laser to overcome the dissipative terms.   
By linearizing around the zero solution, we obtain a characteristic equation
whose eigenvalues determine the stability of the steady state.    
The actual form of the characteristic equation is not shown here due to
a large number of terms, resulting from a $6 \times 6$ matrix
corresponding to a $6$-dimensional system obtained 
when Eqns. (\ref{eq:lasers1})-(\ref{eq:lasers}) are linearized. 
The delay term
in the coupling introduces an exponential term $\exp \left(-2 \tau \lambda
\right)$ in the characteristic equation, where $\lambda$ is a complex 
eigenvalue.
This transcendental function of eigenvalues in the characteristic equation is
typical of delay differential equation systems and can result in an 
infinite number of roots.  This
is in  contrast to systems without delays, where the number of
eigenvalues (and hence roots of the characteristic equation)
corresponds to the number of variables in the system.

\begin{figure}[h]
\hspace*{-1 cm}
{
\epsfxsize=6in
\epsffile{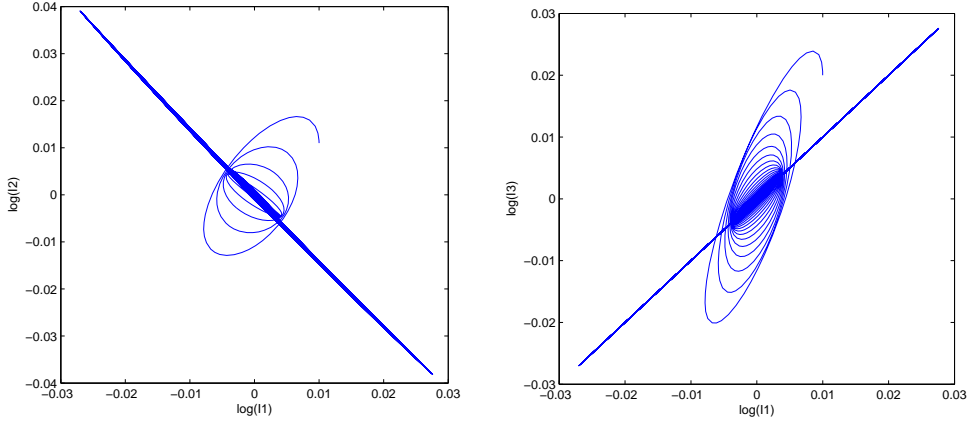}
}
\caption{Synchronization of semiconductor lasers after the transients die 
out, low amplitude regular motion.  
Left: Intensity of Laser 2 vs. Laser 1.  The inner laser
is $180$ degrees out-of-phase with the outer laser, as indicated
by the negative slope of the line. Right:
Laser 3 vs. Laser 1.  Straight line with $+1$ slope indicates 
complete synchronization of outer lasers.
$\tau=\pi/2$, $\epsilon=\sqrt{0.001}$, $\delta_1 = \delta_2 = \epsilon
\sqrt{2.1}$, $a_1 = a_2 =2$, $b_1 = b_2 =1$. }
\label{fig:halfpi}
\end{figure}

As the coupling strengths $\delta_1$ and $\delta_2$ are increased, the system
undergoes a Hopf bifucation where the real parts of the eigenvalues change from negative 
to positive, leading to an onset of oscillations.  
To identify a point of bifurcation, we set the real part of $\lambda$ to
zero: $\lambda = i \omega$.  The transcendental function separates into
real and imaginary parts:  $\exp \left(-2 i \tau \omega \right) = \cos
\left(2 \tau \omega \right) - i \sin \left(2 \tau \omega \right)$.
From the characteristic equation, we now obtain two equations, with
the real part given by 
\begin{equation}
-\omega^6 + A_r \omega^4 + B_r \omega^2 + C_r \omega + D_r = 0 ,
\label{eq:real}
\end{equation}
where $A_r = 2 + 2 \epsilon^2 \beta^2 a_1 a_2 + \epsilon^2 a_1^2 + 
\beta^2$,
$B_r = \beta^2 \delta^2 \cos \left(2 \omega \tau \right) - 1 - 2 \beta^2 - 
2 \epsilon^2 \beta^2 a_1 a_2 - \epsilon^2 a_1^2 \beta^2$; $C_r = -\epsilon
a_1 \beta^2 \delta^2 \sin \left(2 \omega \tau \right)$, $D_r = \beta^2 
\left(1 - \delta^2 \cos \left(2 \omega \tau \right) \right)$, and 
$\delta^2 = 2 \delta_1 \delta_2$. 
The imaginary part of the equation results in
\begin{equation}
A_i \omega^5 + B_i \omega^3 + D_i \omega^2 + F_i \omega + G_i = 0 ,
\label{eq:imaginary}
\end{equation}
where $A_i = 2 \epsilon a_1 + \epsilon \beta^2 a_2$, $B_i = -2 \epsilon
\beta^2 a_2 - 2 \epsilon a_1 - 2 \beta^2 \epsilon a_1 - \epsilon^3 a_1^2
\beta^2 a_2$, $D_i = -\beta^2 \delta^2 \sin  \left(2 \omega \tau \right)$,
$F_i = \beta^2 \epsilon \left(2 a_1 + a_2 \right) - \beta^2 \delta^2 \epsilon
a_1 \cos \left(2 \omega \tau \right)$, and 
$G_i = \beta^2 \delta^2 \sin  \left(2 \omega \tau \right)$.
For no detuning ($\beta=1$), we assume $\omega=1$.
This assumption is
not always valid but is justified for certain values of the delay, as we
shall see shortly.  Solving Eqns.~(\ref{eq:real}) and 
(\ref{eq:imaginary})
for $\beta=1$, and $\omega=1$, we obtain the bifurcation equation:
\begin{equation}
\delta_1 \delta_2 \cos \left(2 \tau \right)  = - \frac{\epsilon^2 a_1 
a_2}{2} .
\label{eq:bifur1}
\end{equation}
It follows that for  values of the delay 
given by $\tau = \left(\pi/2\right) + n \pi$,
where $n$ is an integer, the onset of oscillations occurs when 
\begin{equation}
\delta_1 \delta_2 >  \frac{\epsilon^2 a_1 a_2}{2} .
\label{eq:bifur}
\end{equation}
In this case, the outer lasers are $180$ degrees
out of phase with the middle laser, and synchronized with each other,
after the transients die out.
This effect is plotted in Figure \ref{fig:halfpi}, where the slope of the
line for the log intensity plots of the middle laser 
vs. the outer laser is negative, 
indicating that the two lasers are $180$ degrees out of phase with
each other.  At the same time, the two outer lasers fall on a straight
line of slope $1$ in the log intensity plot,
indicating complete synchronization.  The circular dynamics around the
straight line show
the slow die out of transients as the amplitude of oscillation slowly
increases from its initial conditions.  
It can be seen by substituting the parameters given in 
Fig.~\ref{fig:halfpi} into Eq. (\ref{eq:bifur}) that the coupling
strengths are just above the bifurcation value, leading to low amplitude
regular oscillations.  This can be contrasted to  much higher amplitude
chaotic oscillations shown in Fig.~\ref{fig:lasers_synch}, which will be
treated in the following section.  

The oscillations are regular at low amplitudes, since the dynamics are
approximated by three coupled simple harmonic oscillators due to the dominance
of linear terms when the equations are linearized about the steady state.
The bifurcation condition in Eq. (\ref{eq:bifur}) can be understood as the
point where the coupling strength, $\delta_1 \delta_2$, 
between the lasers is strong enough to
overcome the internal dissipation, which is proportional to $\epsilon a_1$ and
$\epsilon a_2$ for the outer and the inner lasers, respectively.

\section{Theory of Chaotic Synchronization of Semiconductor Lasers}
\begin{figure}[h]
\hspace*{-1 cm}
{
\epsfxsize=6in
\epsffile{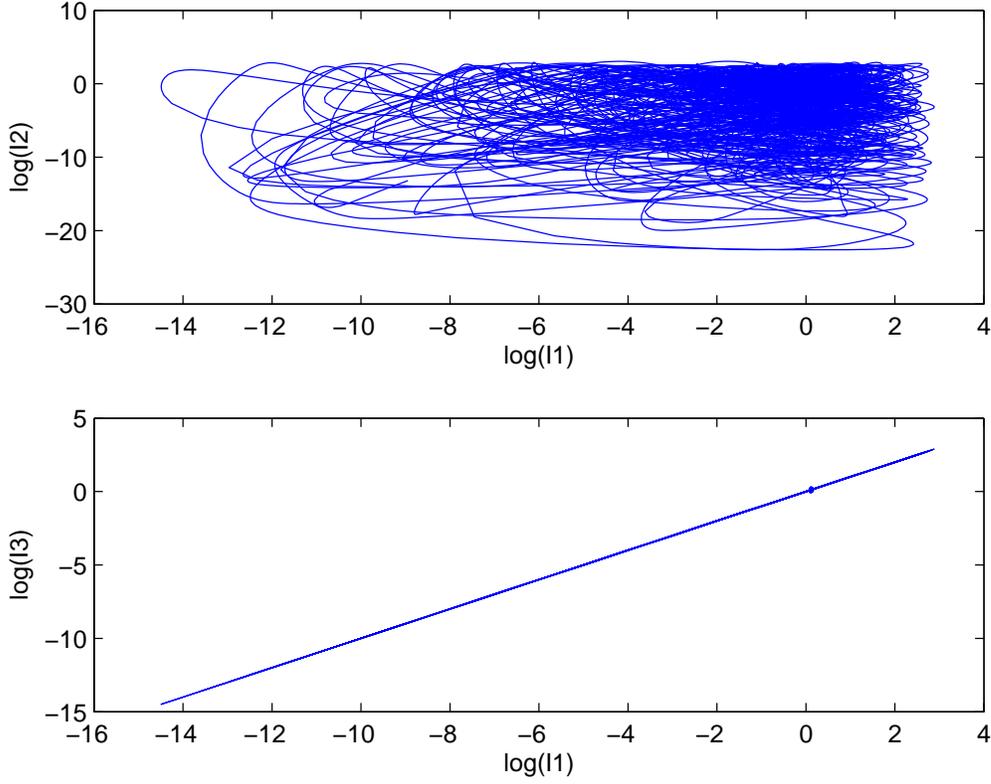}
}
\caption{Chaotic semiconductor lasers.  Top: Intensity of Laser 2 vs. 
Laser 1.  Bottom:
Laser 3 vs. Laser 1.  Straight line indicates 
complete chaotic synchronization of outer lasers.
$\tau=70$, $\epsilon=\sqrt{0.001}$, $\delta_1 = \delta_2 = 7.5 \epsilon$. }
\label{fig:lasers_synch}
\end{figure}

\begin{figure}[h]
\hspace*{-1 cm}
{
\epsfxsize=6in
\epsffile{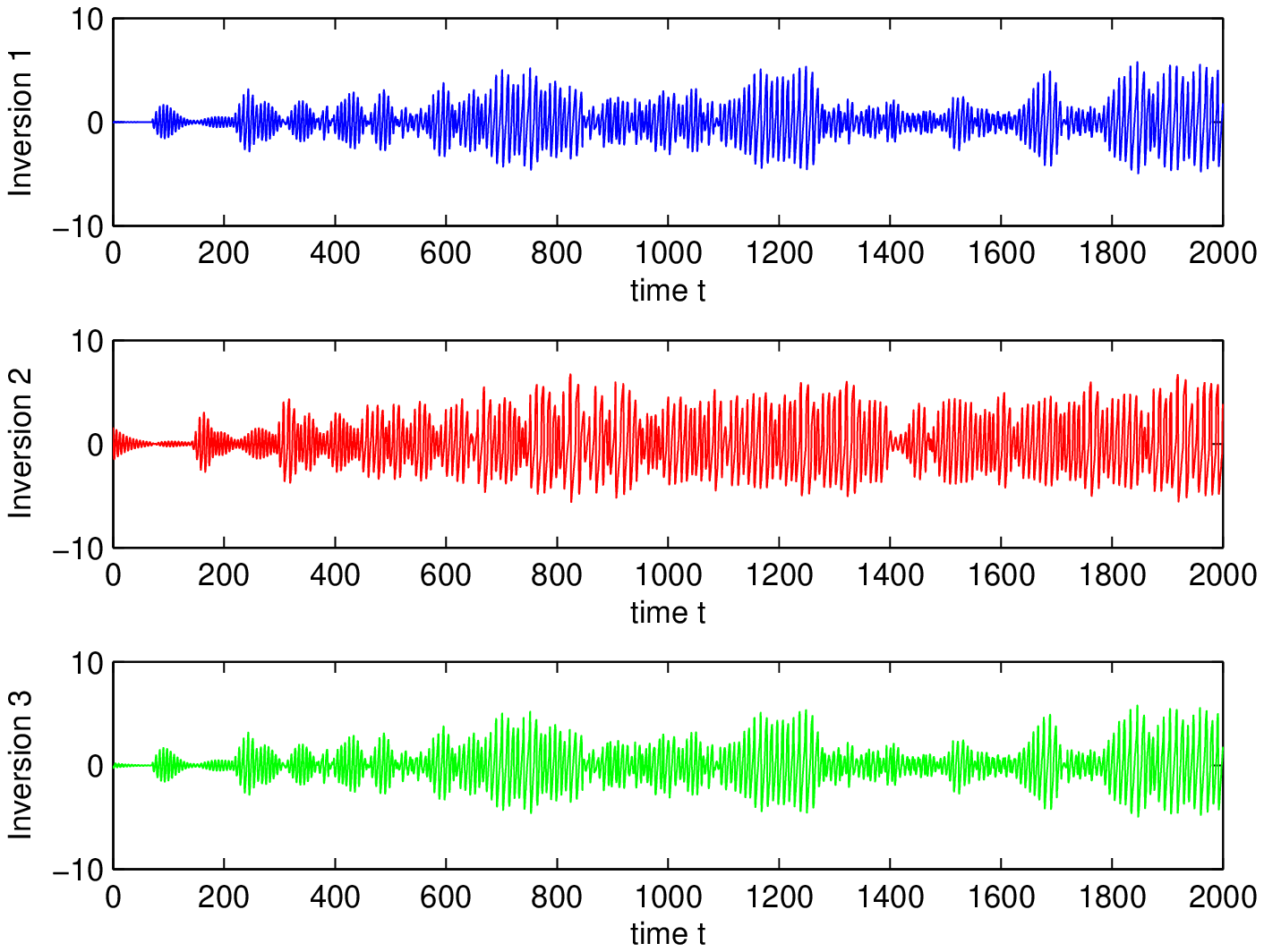}
}
\caption{Inversion of the three semiconductor lasers as a function of 
time, showing
chaotic oscillations.  All parameters the same as in 
Fig. \ref{fig:lasers_synch}.  }
\label{fig:Inv}
\end{figure}

As the coupling strength between semiconductor lasers
is increased further and the amplitude of oscillation
grows, nonlinearities become important and the oscillations become chaotic. 
Chaotic oscillations are shown in Fig.~\ref{fig:Inv} where the inversion,
$x$, is plotted as a function of time.  The coupling strengths in
Fig.~\ref{fig:Inv} are well above the bifurcation value derived in the 
previous section, leading to relatively high amplitudes of oscillation.  
The system described by Eqs. (\ref{eq:lasers1})-(\ref{eq:lasers}) 
shows complete chaotic synchronization
of outer lasers over a wide range of parameters. Figure 
\ref{fig:lasers_synch} shows that while the outer lasers can
become completely synchronized, there may be no apparent correlation between
the middle and the outer lasers.

To understand why chaotic synchronization
occurs without direct contact between the outer lasers and in the presence of
significant delays, we need to consider the basic properties of the system.
Equations (\ref{eq:lasers1})-(\ref{eq:lasers}) can be rewritten in a more 
general form as 
\begin{equation}
\frac{d \mathbf{z_1}}{dt} =  \mathbf{F}\left[\mathbf{z_1}(t)\right] 
+ \delta_1 \cdot \mathbf{G} \left[\mathbf{z_2} \left(t-\tau \right) 
\right]
\label{eq:z1}
\end{equation}
\begin{equation}
\frac{d \mathbf{z_2} }{dt} 
=  \mathbf{\tilde F} \left[\mathbf{z_2}(t)\right] 
+ \delta_2 \cdot \mathbf{\tilde G} \left[\mathbf{z_1} \left(t-\tau 
\right),
\mathbf{z_3} \left(t-\tau \right) \right]
\label{eq:z2}
\end{equation}
\begin{equation}
\frac{d \mathbf{z_3}}{dt} = \mathbf{F} \left[\mathbf{z_3}(t)\right] 
+ \delta_1 \cdot \mathbf{G}\left[\mathbf{z_2} \left(t-\tau \right) \right] 
,
\label{eq:z}
\end{equation}
where $\mathbf{z_i} = \{x_i, y_i\}$.
Due to internal symmetry of the system, there exists an
identical solution for the outer subsystems: 
$\mathbf{z_1}(t) = \mathbf{z_3}(t) \equiv \{X(t), Y(t)\}$.  Whether this
symmetric solution is stable determines whether the lasers will become
synchronized.

Before studying the stability of the synchronized state by
linearizing about the $\{X(t), Y(t)\}$ solution, let us consider a 
qualitative explanation for why complete chaotic synchronization would 
occur in the presence of long delays.  For $\delta_2 =0$, in Eq. (2), 
the dynamics of $\mathbf{z_{1,3}}$ becomes
that of a driven system, with $\mathbf{z_2}$ 
acting as the driver.  Then, the synchronized
dynamics correspond to generalized synchronization \cite{RulkovSTA95}, 
whereby 
the driven subsystem becomes a function of the driver.
While the exact form of the function between the driver and the driven systems
can be rather complicated and 
difficult to obtain, its existence can be inferred from the
synchronization of identical systems when started from different initial
conditions but exposed to the same drive.  This method of detecting
generalized synchronization using identical driven systems is known as the
auxiliary systems approach \cite{AbarbanelRS96}.  
In order for the driven subsystems, $\mathbf{z_{1,3}}$, to
become synchronized, their dependence on initial
conditions has to ``wash out'' as a function of time.  This
independence of later dynamics on initial conditions is necessary
for synchronization.  Otherwise systems that have
different initial conditions will never settle into the same trajectory,
which is necessary for complete synchrony.
This ``washing out'' of initial conditions is provided
by the dissipation in the system, which must therefore be
present in either the coupling term or in the uncoupled dynamics of the
system itself.  For the case of semiconductor lasers that we are considering, 
this dissipation comes from the internal 
dissipation, $\epsilon$, in the lasers themselves.  As was discussed in the
previous section, due to this dissipation the dynamics would spiral down
to  $\mathbf{z_{1,3}}=0$ in the absence of any coupling.  
As will be shown
shortly, this dissipation $\epsilon$ plays an important role in determining
the Lyapunov exponents transverse to the synchronized state.  

In addition to helping understand the unidirectionally driven case, this 
idea of generalized synchronization between the driver and the driven
systems leading to complete synchronization of identical driven systems is
also useful in understanding complete chaotic
synchronization of mutually coupled lasers with long delays \cite{LandsmanS07}.  
In mutually coupled systems (i.e., $\delta_2 \neq 0$), 
the dynamics of $\mathbf{z_2}$ 
are affected by $\mathbf{z_1}$ and $\mathbf{z_3}$.  In this case, the
synchronized state, $\{X(t), Y(t)\}$, will depend on the 
initial conditions of all of the
three subsystems, $\{\mathbf{z_1}, \mathbf{z_2}, \mathbf{z_3}\}$, 
so that $\{X(t), Y(t)\}$ can not be the result
of generalized synchronization, in a strict sense.  However, it takes a
time interval of $2 \tau$ for any change in the dynamics of systems 
$\mathbf{z_{1,3}}$ 
to affect the trajectory of these systems via mutual coupling.  
During this time interval of length $2 \tau$, $\mathbf{z_{1,3}}$ 
can be viewed as driven by $\mathbf{z_2}$, since the signal 
$\mathbf{z_{1,3}}$  receives during that time interval is not affected
by its dynamics on that interval.
This idea that the dynamics of the outer lasers can be viewed as driven
by the signal from the middle laser on the time interval $2 \tau$ will be
very useful when we linearize the dynamics about the synchronized state.
Since dynamics transverse to the synchronized state will not affect the
synchronous solution, $\{X(t), Y(t)\}$, over twice the coupling delay, we can 
separate the
variables and obtain an analytic solution which is valid over that interval.  

To linearize Eqns. (\ref{eq:lasers1}) and (\ref{eq:lasers}) around 
the synchronized state, $\{X(t), Y(t)\}$, we introduce new variables 
defined as: 
$\triangle x_{1,3}(t) = x_{1,3}(t) - X(t)$ and 
$\triangle y_{1,3}(t) = y_{1,3}(t) - Y(t)$.  
If the conditional Lyapunov exponents calculated with respect to perturbation
out of the synchronization manifold are all negative, then the outer lasers are
synchronized.  Calculating Lyapunov exponents is in
general complicated due to the presence of time-delays in the equations.  
The coupling term containing delays, however, drops out of the equations
if we take the difference of the outer variables: 
$\triangle y=y_1-y_3; \triangle x = x_1 - x_3$.  
For simplicity, let us assume
that we only perturb one of the outer lasers from the synchronous
state.  This will not affect the generality of the result, but allows us 
to identify the ``synchronized state'' with the dynamics of the other 
unperturbed laser, so that if $y_3=Y(t)$ and $x_3=X(t)$, then 
$\triangle y= \triangle y_{1,3}(t) = y_1-Y(t)$ 
and $\triangle x = \triangle x_{1,3}(t) = x_1 - X(t)$.  Using notation of
Eqns.~(\ref{eq:z1}) - (\ref{eq:z}), the linearized dynamics 
transverse to the synchronization manifold are given by,
\begin{equation}
\frac{d \mathbf{\triangle z}(t)}{dt} 
= \mathbf{J} \cdot \mathbf{\triangle z}(t)
\label{eq:lin}
\end{equation}
where $\mathbf{\triangle z}(t) = \{\triangle x, \triangle y \}$
and  $\mathbf{J}$ is a $\mathbf{2} \times \mathbf{2}$ Jacobian matrix of partial derivatives evaluated 
at $\mathbf{z} = \{X(t), Y(t)\}$,
\begin{equation}
\mathbf{J}  = \frac{\partial \mathbf{F(z)}}{\partial \mathbf{z}}
\label{eq:Jacobian}
\end{equation}
Applying Eqns. (\ref{eq:lin}) and (\ref{eq:Jacobian}) 
to Eqn. (\ref{eq:lasers1}) (after comparing it to the more general form
of Eqn. (\ref{eq:z1})) we obtain
\begin{equation}
\begin{pmatrix} \dot{\triangle x} \\ \dot{\triangle y} \end{pmatrix} =
\begin{pmatrix} -\epsilon \left( a_1 + b_1 Y(t) \right) 
& -\left( 1+\epsilon b_1 X(t) \right) \\ 1+Y(t) & X(t) \end{pmatrix} \cdot 
\begin{pmatrix} \triangle x(t) \\ \triangle y(t) \end{pmatrix}
\label{eq:matrix}
\end{equation}
Both $1+Y(t)$ and $1+\epsilon b_1 X(t)$ terms in the matrix of
Eqn. (\ref{eq:matrix}) are positive.
The first because $Y(t) > -1$, as follows from Eqns. 
~(\ref{eq:lasers1})-(\ref{eq:lasers}), and the second 
because $|\epsilon b_1 X(t)| < 1$  since $\epsilon \ll
1$. It follows that the cross-terms in the matrix always have opposite
signs, indicating a finite counter-clockwise rotation of a system with
instantaneous frequency given by,
\begin{equation}
\omega(t) = \left( 1+Y(t) \right)^{1/2} \left( 1+\epsilon b_1 X(t)
\right)^{1/2} 
\label{eq:instfreq}
\end{equation}
The angular frequency in the above equation shows that the speed of rotation
of the transverse dynamics varies as a function of time, but is
always non-vanishing and counter-clockwise.  We will come back to this
property shortly in connection to proving the stability of synchronized
state for sufficiently long delays.  

If the dynamics of the outer laser are perturbed from the synchronized
state at some time $t=t_0$, then the perturbation will not affect 
the coefficient matrix in Eqn. (\ref{eq:matrix})
until $t \geq t_0 + 2 \tau$.  It follows that during the time interval 
of $2 \tau$ the dynamics off the synchronization manifold can be 
viewed as driven by the time dependent coefficients: $\{X(t), Y(t)\}$.   
Since on this timescale, the
coefficients in the matrix are independent of the variables $\{\triangle x,
\triangle y \}$, Eqn.~(\ref{eq:matrix}) can be solved over the interval.  
This is much like solving an equation of the form $dh/dt = f(t) h$.  As 
long as
$f(t)$ is only a function of time and independent of $h$, we can easily
obtain a solution: $h = exp \left(\int f(t) dt \right)$.  The situation in
Eqn.~(\ref{eq:matrix}) is similar, but in two dimensions:  as long as
we are looking at the interval of $2 \tau$, the variables $\{X(t), Y(t)\}$ can
be viewed as some functions of time only, since they are independent of
$\{\triangle x, \triangle y \}$ over that interval.  
Thus we can choose any initial condition 
$\{\triangle x_0, \triangle y_0 \}$ at some time $t=t_0$ for a small
perturbation transverse to the synchronized state and solve for the dynamics
on the time interval of $t_0 \leq t \leq t_0 + 2 \tau$.  Of course we still
have not solved for the matrix coefficients in Eqn.~(\ref{eq:matrix}),
since $X(t)$ and $Y(t)$ are solutions of a time-delayed differential
equations and, in general, can not be easily obtained by analytic means.
However, they have special properties, namely $X(t)$, the inversion, is
symmetric about zero, and $Y(t)$, the intensity is always greater than 
$-1$.  As we
shall see shortly, these properties will allow us to make some conclusions
about the stability of synchronized state without actually having to solve
for $\{X(t), Y(t) \}$.   

A two dimensional equation with time-dependent coefficients can be solved
using Abel's formula, which relates the Wronskian of the linearized system
to the trace of the matrix \cite{zill}
\begin{equation}  
W\left(t\right) = \det \begin{vmatrix} \triangle x & \triangle y
\\ \dot{\triangle x} & \dot{\triangle y} \end{vmatrix} 
= W_0 \exp\left( \int^t_{t_0} \{X(s) - \epsilon \left(a_1 + b_1 Y(s) \right) \} \cdot ds \right)
\label{eq:W}
\end{equation}
where $W_0>0$ is a constant that depends on the magnitude of the initial
perturbation: $W_0 = W(t_0)$.
The Wronskian gives the phase-space volume dynamics  of the system 
$\{\triangle x(t), \triangle y(t) \}$.  Equation (\ref{eq:W}) is valid over the
integration interval of twice the delay: $t_0 < t < t_0 + 2 \tau$.
The term multiplied by $\epsilon$ in the exponential is always negative,
since $a_1 > b_1$, and $Y(s) > -1$, leading to the contraction of phase-space
volume.  The inversion term, $X(s)$, on the other hand, is symmetric around
zero, resulting in zero average over the oscillations.  Since the fluctuations
in inversion don't have a prefered direction, their statistical average is
zero.  Thus for sufficiently long delays, 
the $- \int \epsilon \left(a_1 + b_1 Y(s) \right) dt$ term will always
dominate in the exponential of Eqn. (\ref{eq:W}).  This term
is always negative, assuming $a_1 > b_1$, and
monotonically decreasing as a function of delay (which determines the length
of integration), resulting in shrinkage of the
phase-space volume for sufficiently long delays.  

Taking the determinant of the matrix in the above equation, $W(t)$, can
also be written as $W(t) =  |\triangle x \dot{\triangle y} 
- \triangle y \dot{\triangle x}|$.
Substituting for $\dot{\triangle x}$ and $\dot{\triangle y}$
from Eqn.~(\ref{eq:matrix}), we obtain
\begin{equation}
W(t) = |\left(1+Y(t)\right) \cdot \left(\triangle x\right)^2 
+ \left(1+\epsilon b_1 X(t)\right) \cdot \left(\triangle
y\right)^2 + \left[ \epsilon \left( a_1 + b_1 Y(t) \right) + X(t) \right] 
\cdot
\triangle x \triangle y |
\label{eq:Wcontract}
\end{equation}
Since both $1+Y(t)$ and $1+\epsilon b_1 X(t)$ terms in the above equation 
are positive, the terms quadratic in $\triangle x$ and $\triangle y$ 
are positive as well.  This is characteristic of the phase-space
volume of rotating systems.
Due to ever-present finite counter-clockwise rotation in the system
(see discussion following Eqns. (\ref{eq:matrix}) and (\ref{eq:instfreq})), 
if the phase space volume, $W(t)$, is shrinking over several rotations, 
then the distance from synchronized state, 
$r=\sqrt{\left(\triangle x\right)^2 + \left(\triangle y\right)^2}$, has to
shrink as well.
This can be seen in the following way:  suppose we draw a straight
line, given by $\triangle y=c \triangle x$, through the origin
in the phase space plot of  $\{\triangle x, \triangle y\}$, where 
the slope, $c$, is some arbitrary constant.  
Now, suppose we are interested in the value of $W(t)$ whenever this
line is crossed.  To find $W(t)$ at the point of crossing, we can 
substitute $\triangle y=c \triangle x$
into Eqn.~(\ref{eq:Wcontract}),
\begin{equation}
W(t) = \left[1+Y(t) 
+ c \left[\epsilon \left(a_1 + b_1 Y(t)\right) +X(t) \right]
+ c^2 \left(1 + \epsilon b_1 X(t) \right) \right] \triangle
x^2
\label{eq:crossing}
\end{equation}
Since, as explained following 
Eqns. (\ref{eq:matrix}) and (\ref{eq:instfreq}),
there is a non-vanishing counter-clockwise 
finite rotation in the system, this line $\triangle y= c \triangle x$ will 
be crossed at each successive rotation, for any value of the slope, $c$. 
The first factor in the above equation depends on the variables $X(t)$ 
and $Y(t)$, which come from an arbitrarily
chosen interval of $2 \tau$ and therefore do not have any consistant
time-dependent behavior within that interval.  It follows that if $W(t)$
always decreases after a certain period of time, then 
$\{\triangle x, \triangle y \}$ have to decrease along any line
$\triangle y=c \triangle x$ drawn from the origin.  This is just shrinking
of the radius, $r$, or distance transverse to the synchronized state.  
We have thus shown that monotonic shrinking of the phase-space volume
of the perturbed dynamics corresponds to stability of synchronized state.

It remains for us to show that $W(t)$ always shrinks towards the end of
twice the delay time for sufficiently long delays.  We do this by showing
that the integral in Eqn.~(\ref{eq:W}) becomes negative for sufficiently
long integration times.
Since the variable $Y(t)$ is the scaled intensity of the laser, from 
Eqsn.~(\ref{eq:lasers1})-(\ref{eq:lasers}) its minimum possible value is
$-1$.  Thus for $a_1 > b_1$ (a typical case), the contribution of the
dissipation term to the Wronskian is always negative.
The variable $X(t)$, on the other hand, is symmetric about zero, and
thus averages out to zero when integrated over a single period of oscillation.
It follows that $X(s)$ in Eqn.~(\ref{eq:W}) 
averages out to zero if the integral is done over many periods of oscillation,
while the dissipation term, multiplied by $\epsilon$, provides a continuous
negative component.  If that continuous negative component builds up
sufficiently over the integration interval to overcome any fluctuations
in $X(s)$, we then have a continuous shrinking of the phase-space of
perturbed dynamics, which combined with rotation in the two dimensional  
system leads to synchronization.  

We can now understand how synchronization is dependent on delays.  The upper
limit on the integration interval in Eqn.~(\ref{eq:W})
is set at twice the delay time. If the delays are sufficiently
long to overcome fluctuations in $X(t)$, then the phase-space dynamics
transverse to synchronized state shrink.  Depending on the value of 
dissipation, $\epsilon$, the integration time may have to be long 
to consistently get a negative exponent in Eqn.~(\ref{eq:W}).  This is due
to the fact that
$\epsilon \ll 1$, while the fluctuations in  $X(s)$, although zero when
integrated over a period, will introduce fluctuations of order one into
the integral.  The upper limit on the positive
fluctuation in Eqn.~(\ref{eq:W}) is given by $T |X(t)|_{max}/2$, where 
$|X(t)|_{max}$ is the maximum value of the intensity and $T$ is the
corresponding period of oscillation.  This is
the maximum value that the integral of any oscillation symmetric about 
zero of period $T$ and amplitude $|X(t)|_{max}$ can reach, leading to
\begin{equation}
\int^{t_0+2 \tau}_{t_0} X(s) ds < T |X(t)|_{max}/2
\label{eq:t0}
\end{equation}
The dissipative term in the exponential of Eqn.~(\ref{eq:W}) is
equivalent to
\begin{equation}
\int^{t_0+2 \tau}_{t_0} \epsilon \left(a_1 + b_1 Y(s) \right) = 2 \tau 
\epsilon  \left(a_1 + b_1 \bar{Y} \right)
\label{eq:dissadd}
\end{equation}
where $\bar{Y}$ is the averaged intensity.  Applying Eqns.~(\ref{eq:t0}) 
and
(\ref{eq:dissadd}) to Eqn.~(\ref{eq:W}), we obtain a condition for the 
shrinking
of transverse phase-space volume toward the end of the interval of twice the
delay time,
\begin{equation}
4 \epsilon  \left(a_1 + b_1 \bar{Y} \right) \left( \frac{\tau}{T} \right) > |X(t)|_{max}
\label{eq:compare2}
\end{equation}
From the above equation, it is clear that the ratio of the delay to the
period of oscillation, $\tau/T$, plays an important role in synchronization.  
Since the dissipation, $\epsilon$, is
small, we need rather long delays, compared to the period of oscillation,
to guarantee the stability of synchronized state.  
When the fluctuations in intensity are
sufficiently small so that the period $T$ is approximated well by $2 \pi$
in our scaled equations, Eqn.~(\ref{eq:compare2}) reduces to
$2 \epsilon  \tau \left(a_1 + b_1 \bar{Y} \right)  > \pi |X(t)|_{max} .$

Since the solution in Eqn.~(\ref{eq:W}) is no longer valid for integration
times longer than twice the delay, we need to consider what happens at
the end of that interval.  At the beginning of the new interval
at $t=t_0 + 2 \tau$, our
synchronized state has been affected by the dynamics of $\{\triangle x(t),
\triangle y(t) \}$ over the previous interval.  Let us call this new
synchronized state $\{X(t) \prime, Y(t) \prime \}$.  This is of course
the same as saying that the perturbation of one of the outer lasers has
finally reached the other, after a time of $2 \tau$, and affected
the ``synchronized state.''  What we are really
interested in is the evolution of perturbation $\{\triangle x(t),
\triangle y(t) \}$ from the altered dynamics, $\{X(t) \prime, Y(t) \prime \}$,
of the other laser.  This is because we are interested in whether
the outer lasers will become synchronized 
(even if the ``synchronized state'' changes),
not in whether they will come back to the same synchronized state that 
would have existed if the perturbation never happened.
In addition, we have to consider that
the time delay terms in the original Eqns.~(\ref{eq:lasers1}) -
(\ref{eq:lasers}) are affected by a perturbation after a time of $2 \tau$,
so that we can no longer get the delay terms to drop out of the equations
by linearizing around the same synchronized solution that would 
have existed in the absence of perturbations.  Hence this is not quite 
the same as using linearization to find the
divergence of two nearby trajectories in phase space that are governed
by the same equation but have different initial conditions.  In our case,
the two nearby trajectories affect each other and thus can not be 
considered
to evolve independently.  In other words, the transverse dynamics affect the
synchronized state dynamics at a later time, so we can not just linearize  
around the synchronized state the way we would linearize to find the
divergence of two nearby independent
trajectories.   

The perturbation
from synchronized state at the beginning of the new time interval is given
by $\{\triangle x(t_0 + 2 \tau), \triangle y(t_0 + 2 \tau) \}$.  
Since the transverse 
dynamics are again independent over the period of $2 \tau$ from the
synchronized dynamics  $\{ X(t) \prime, Y(t) \prime\}$, we can again
apply Abel's formula formula over that period, with the initial condition
of $\{\triangle x(t_0 + 2 \tau), \triangle y(t_0 + 2 \tau) \}$.  However, 
$\{X(t) \prime, Y(t) \prime \}$ have the same properties as 
the synchronized solution before,
namely, $X(t) \prime$ is symmetric about zero and $Y(t) \prime > -1$.
We have already shown that given these properties, the distance from
synchronized state,  $r=\sqrt{\left(\triangle x\right)^2 + \left(\triangle
  y\right)^2}$, will shrink towards the end of twice the delay time.
Applying the same arguement to the next interval, we can see that
$r(t_0) > r(t_0 + 2 \tau) > r(t_0 + 4 \tau) ...$ ad infinitum.  We have
thus proved that the synchronized state of the outer lasers is stable
for sufficiently long delays. 

We have shown that the synchronized state is stable for sufficiently
long delays.  The stability of the synchronized state indicates that
all the transverse Lyapunov exponents are negative.  A negative Lyapunov 
exponent
sum corresponds to the contraction of the phase-space volume.  In fact
there is a simple relationship beween the two, given by:
\begin{equation}
\lambda_1 + \lambda_2 = \lim_{\triangle t\to\infty} \frac{1}{\triangle t} 
\ln \left(W(t_0 + \triangle t)/W_0\right),
\label{eq:LyapSum}
\end{equation}
where $\lambda_1 + \lambda_2$ is the sum of transverse Lyapunov exponents.
Based on Eqn.~(\ref{eq:W}) and prior discussion, it is clear that
for sufficiently long delays, the phase space volume over each interval
of $2 \tau$ contracts by a factor of approximately
$ \exp \left(- 2 \tau \epsilon \left(a_1 + b_1 \bar Y\right) \right)$,
so that over $n$ intervals, the phase-space volume is approximated by
\begin{equation}  
\frac{W\left(t_0 + 2 \tau n\right)}{W_0} 
\approx \exp \left(- 2 \tau n \epsilon \left(a_1 + b_1 \bar Y\right) \right)
\label{eq:Wn}
\end{equation}
Taking $\triangle t = 2 \tau n$ and $n \rightarrow \infty$ 
for infinite times, we
obtain after substituting Eqn.~(\ref{eq:Wn}) into Eq.~(\ref{eq:LyapSum}),
\begin{equation}
\lambda_1 + \lambda_2 \approx -\epsilon \left(a_1 + b_1 \bar{Y} \right)
\label{eq:LyapLasers}
\end{equation} 
The sum of 
Lyapunov exponents shows a negative linear dependence on
dissipation whenever the lasers synchronize.  This intimate connection
between the Lyapunov exponents and dissipation is not accidental, since
negative Lyapunov exponents determine how quickly the perturbed trajectory
converges to the synchronized state, and the dissipation determines how
quickly any differences in initial conditions of the outer lasers ``wash 
out,'' leading to synchronization.

\begin{figure}[h]
\hspace*{-1.5 cm}
{
\epsfxsize=5in
\epsffile{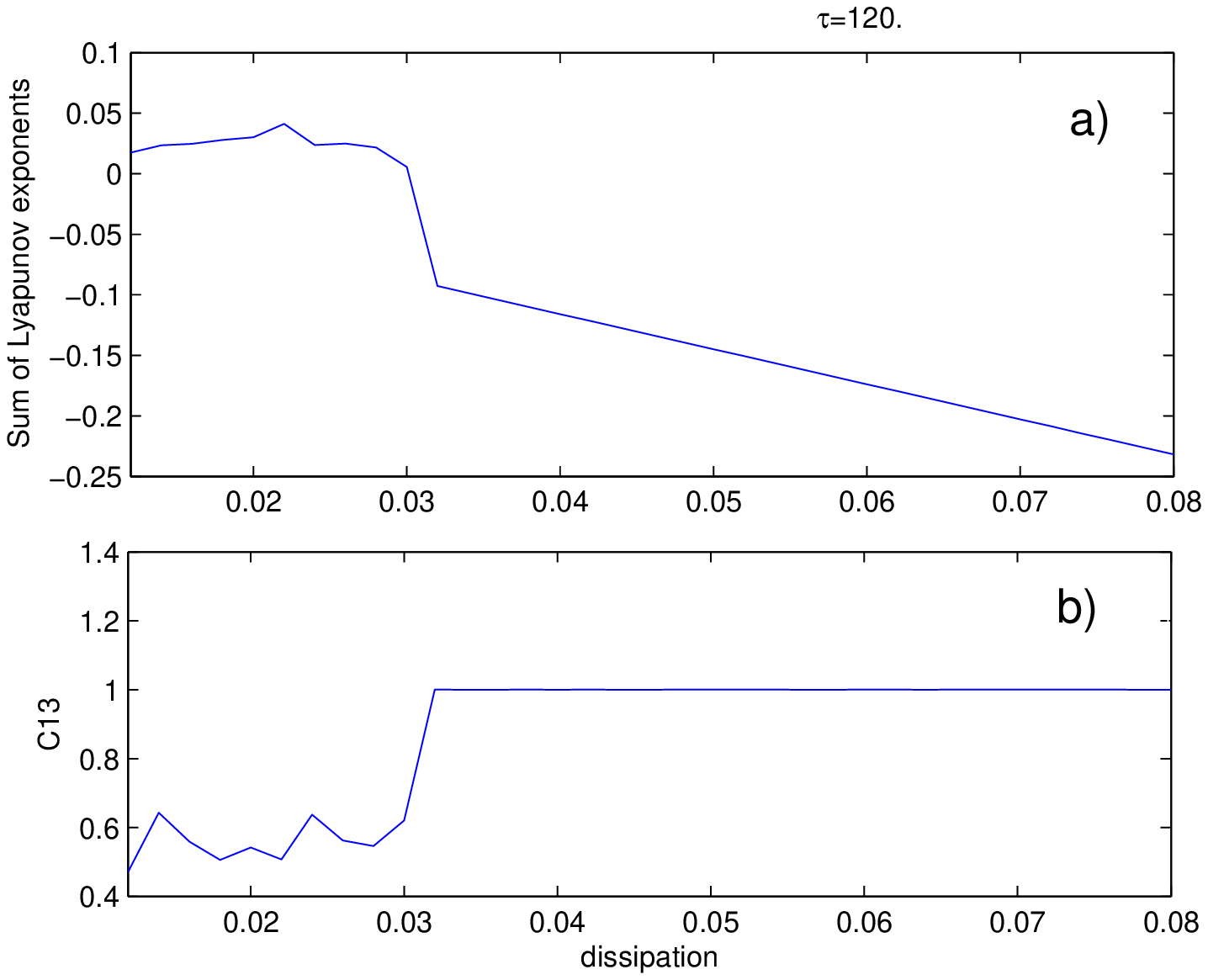}
}
{
\hspace*{-1.3 cm}
\epsfxsize=5in
\epsffile{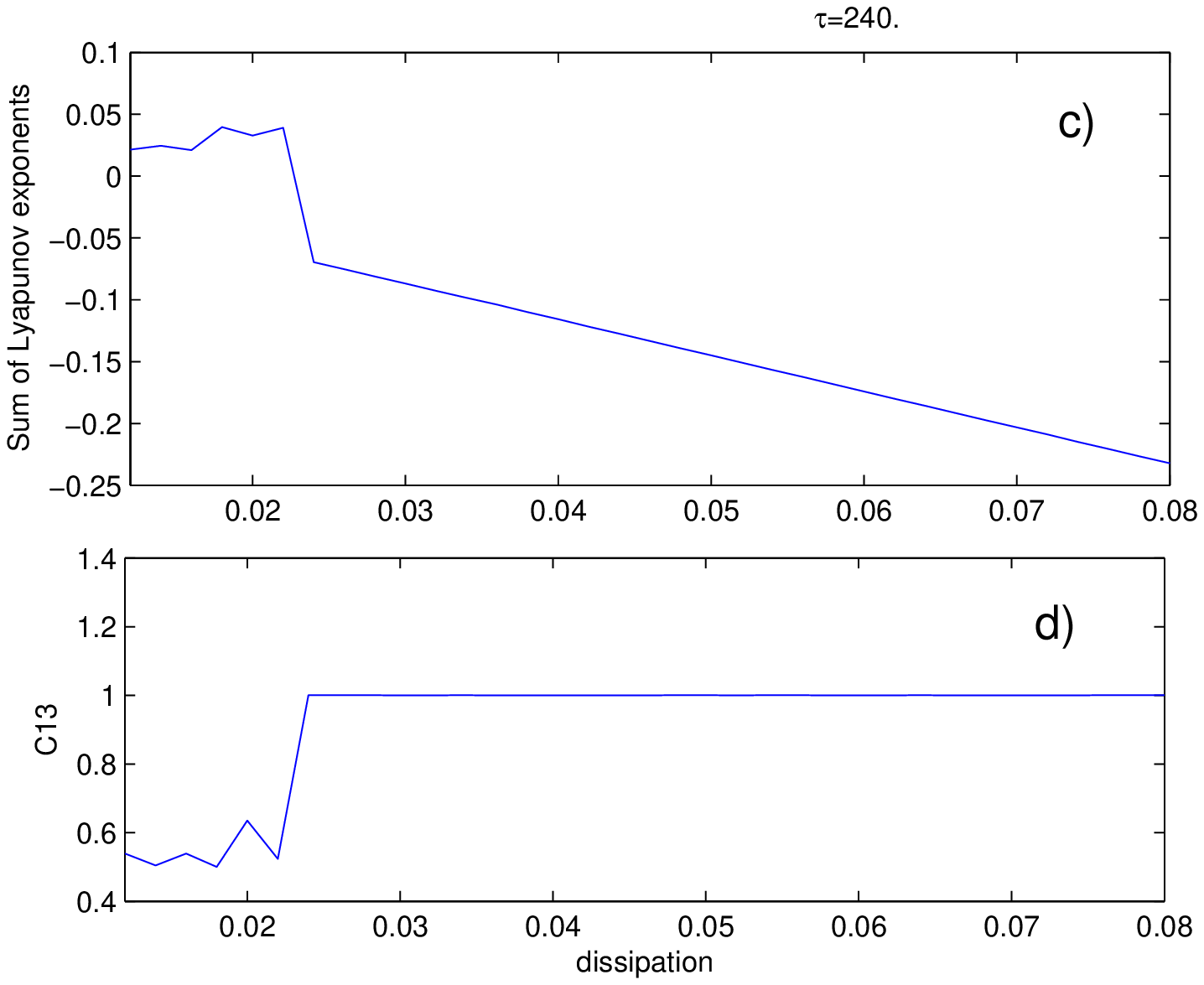}
}
\caption{a) Sum of Lyapunov exponents as a function of dissipation, $\epsilon$,
for $\tau=120$.  b) Corresponding
  correlations between outer lasers, $\tau=120$.  
c) Sum of Lyapunov exponents vs. $\epsilon$,  
for $\tau=240$.  d)  Corresponding
  correlations between outer lasers, $\tau=240$. 
In all cases,
$a_1=a_2=2$, $b_1=b_2=1$, $\delta_1 = \delta_2 = 0.2$, $\beta=0.5$.}
\label{fig:SumLyapEp}
\end{figure} 
Figure \ref{fig:SumLyapEp} shows the numerically computed sum of
Lyapunov exponents, and corresponding correlations of the outer lasers,
as a function of dissipation, $\epsilon$, for 
two values of the delay, $\tau=120$ and $\tau=240$.  
The fluctuations in the sum of Lyapunov exponents
correspond well to the fluctuations in the correlation function of the outer
lasers, with desynchronization when the Lyapunov sum increases above 
zero.  As might be expected from
Eqn.~(\ref{eq:compare2}), longer delays mean synchronization at lower 
values of
dissipation, since the dissipation term in the exponential in
Eqn.~(\ref{eq:W})  dominates for 
sufficiently long delays. 

\begin{figure}[h]
\hspace*{-1 cm}
{
\epsfxsize=5in
\epsffile{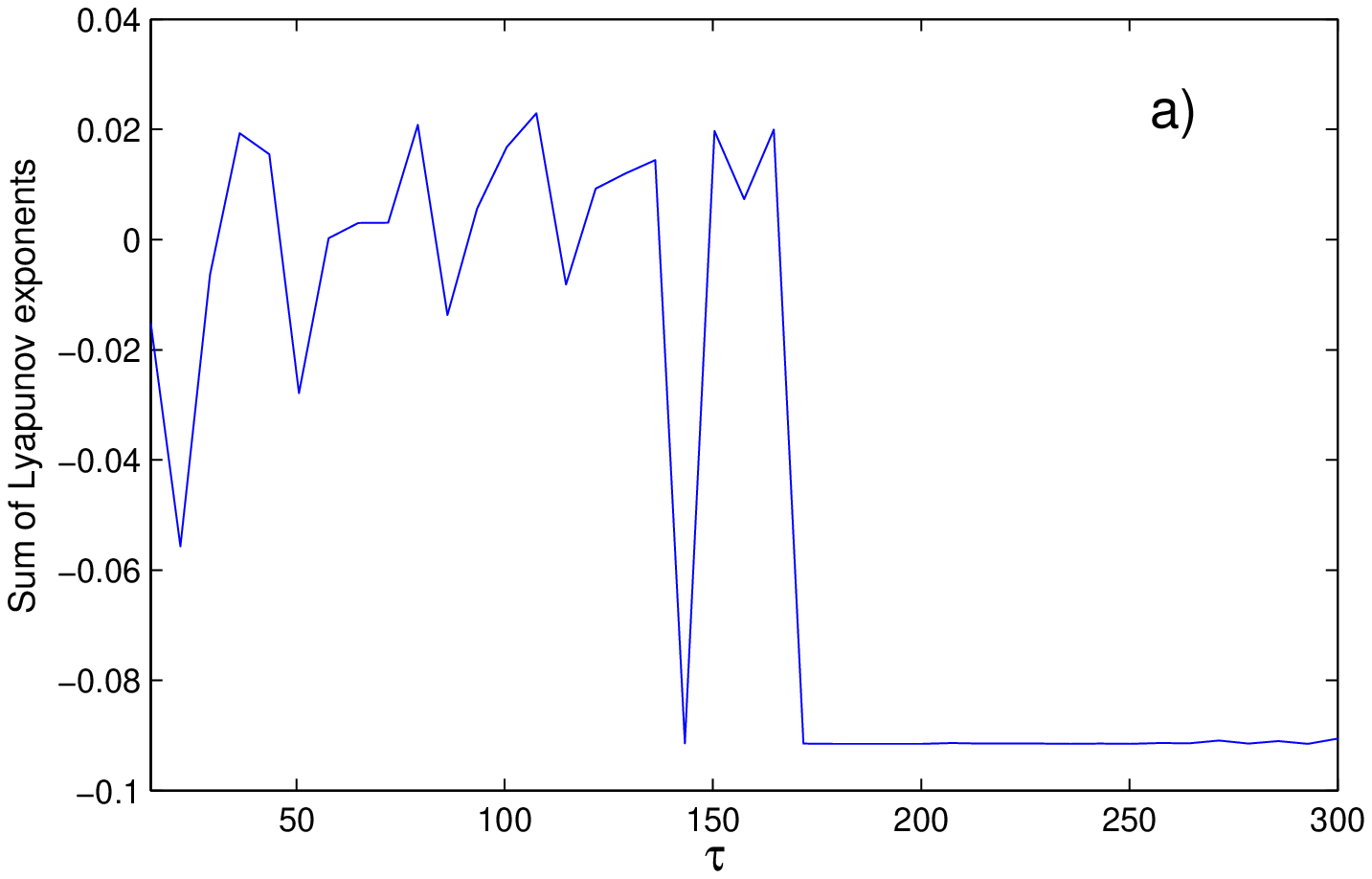}
}
{
\hspace*{-0.25 cm}
\epsfxsize=5in
\epsffile{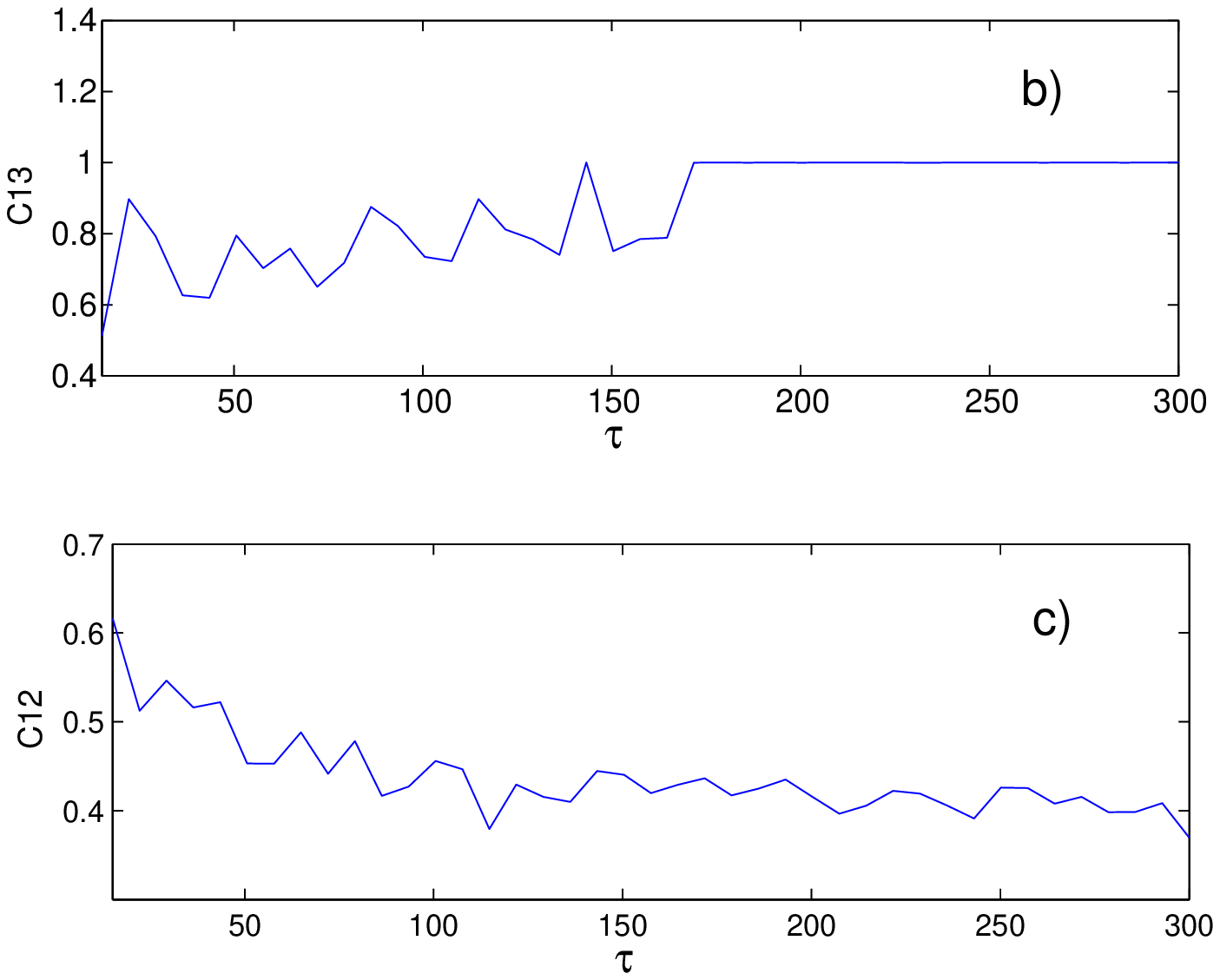}
}
\caption{a) Numerically computed sum of Lyapunov exponents as 
a function of delay, $\tau$.  b) Corresponding correlations of outer lasers.
c) Correlations of the middle and outer lasers, shifted by
the delay time to maximize correlations.  
$\epsilon = \sqrt{0.001}$, $\delta_1=\delta_2=7.5 \epsilon$, $\beta=0.5$.}
\label{fig:SumLyapDelay}
\end{figure}

Figure  \ref{fig:SumLyapDelay}a
shows the sum of Lyapunov exponents as a function of delay.  
The Lyapunov exponents are negative for all $\tau > 170$, (corresponding 
to about $60$ ns) resulting in
complete synchronization of the outer lasers, as shown in 
Fig.~\ref{fig:SumLyapDelay}b.  
At the same time, the outer lasers are not synchronized with the
center one, Fig.~\ref{fig:SumLyapDelay}c.  The fluctuations in 
correlations
of the outer lasers match well the fluctuations in the Lyapunov sum, with
correlations increasing whenever the Lyapunov sum decreases.  
Figure  \ref{fig:SumLyapDelay} agrees well with the analysis in this section, 
since sufficiently long delays are needed for the Lyapunov exponents to become
negative, leading to synchronization.  After the onset of synchronization,
Eqn.~(\ref{eq:LyapLasers}) becomes valid, so the Lyapunov sum
becomes independent of delays.  This is confirmed by the straight horizontal
line in the figure, after the outer lasers synchronize. The degree of
synchrony is given by the correlation function. 

\begin{figure}[h]
\hspace*{-1 cm}
{
\epsfxsize=6in
\epsffile{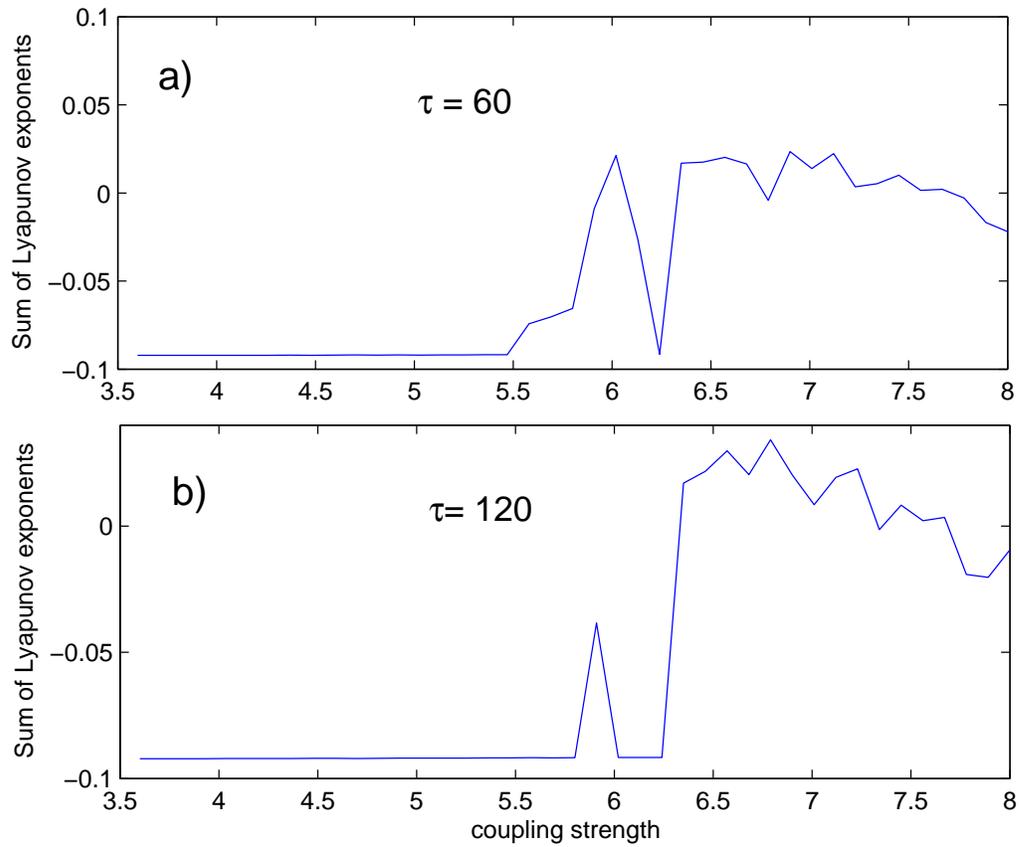}
}
\caption{a) Sum of Lyapunov exponents as 
a function of coupling strength, $\delta_1=\delta_2$, for $\tau=60$.  
b) $\tau=120$. 
$\epsilon = \sqrt{0.001}$, $\beta=0.5$.}
\label{fig:SumLyapCoupling}
\end{figure}

\begin{figure}[h]
\hspace*{-1 cm}
{
\epsfxsize=5in
\epsffile{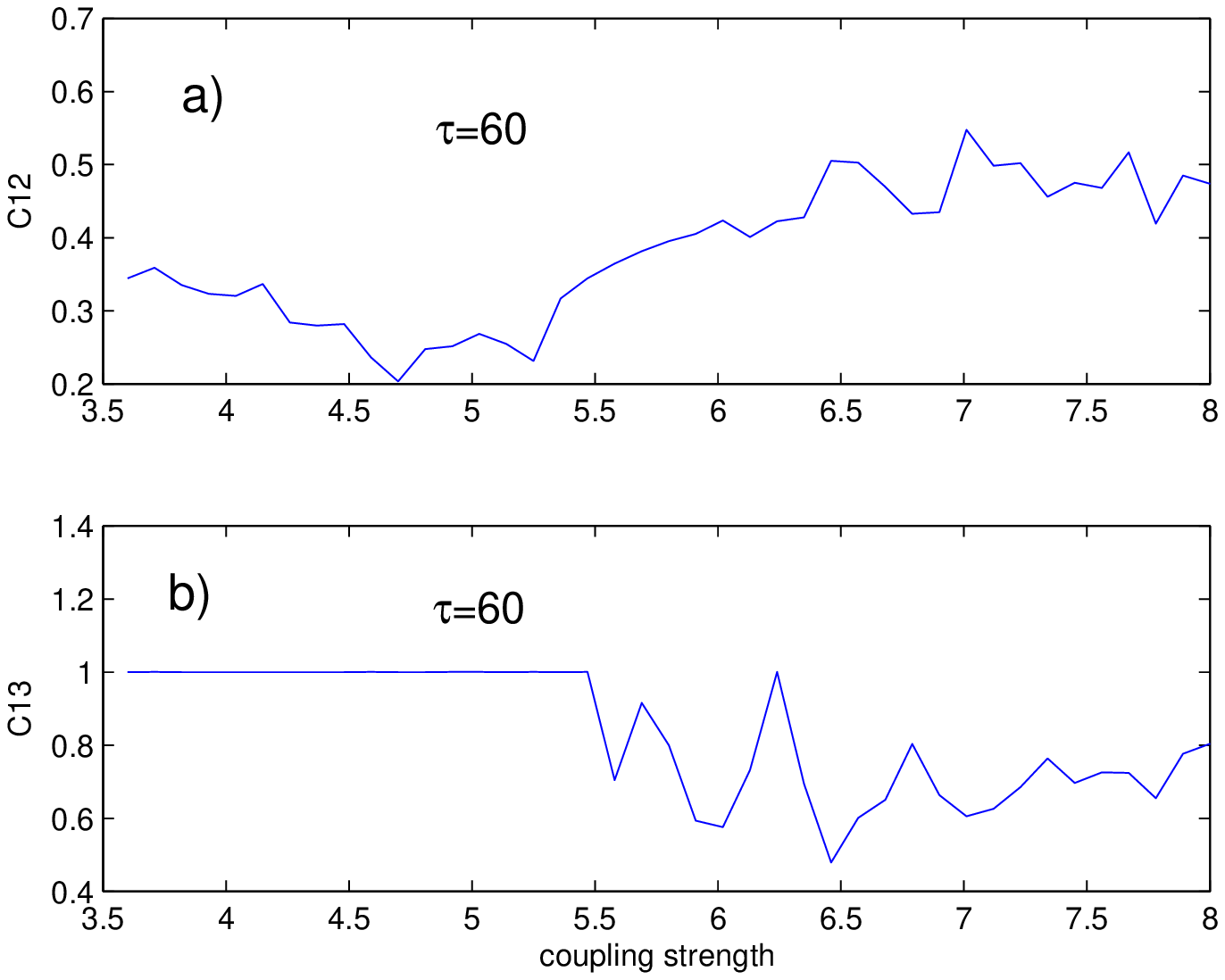}
}
{
\hspace*{-0.7 cm}
\epsfxsize=5in
\epsffile{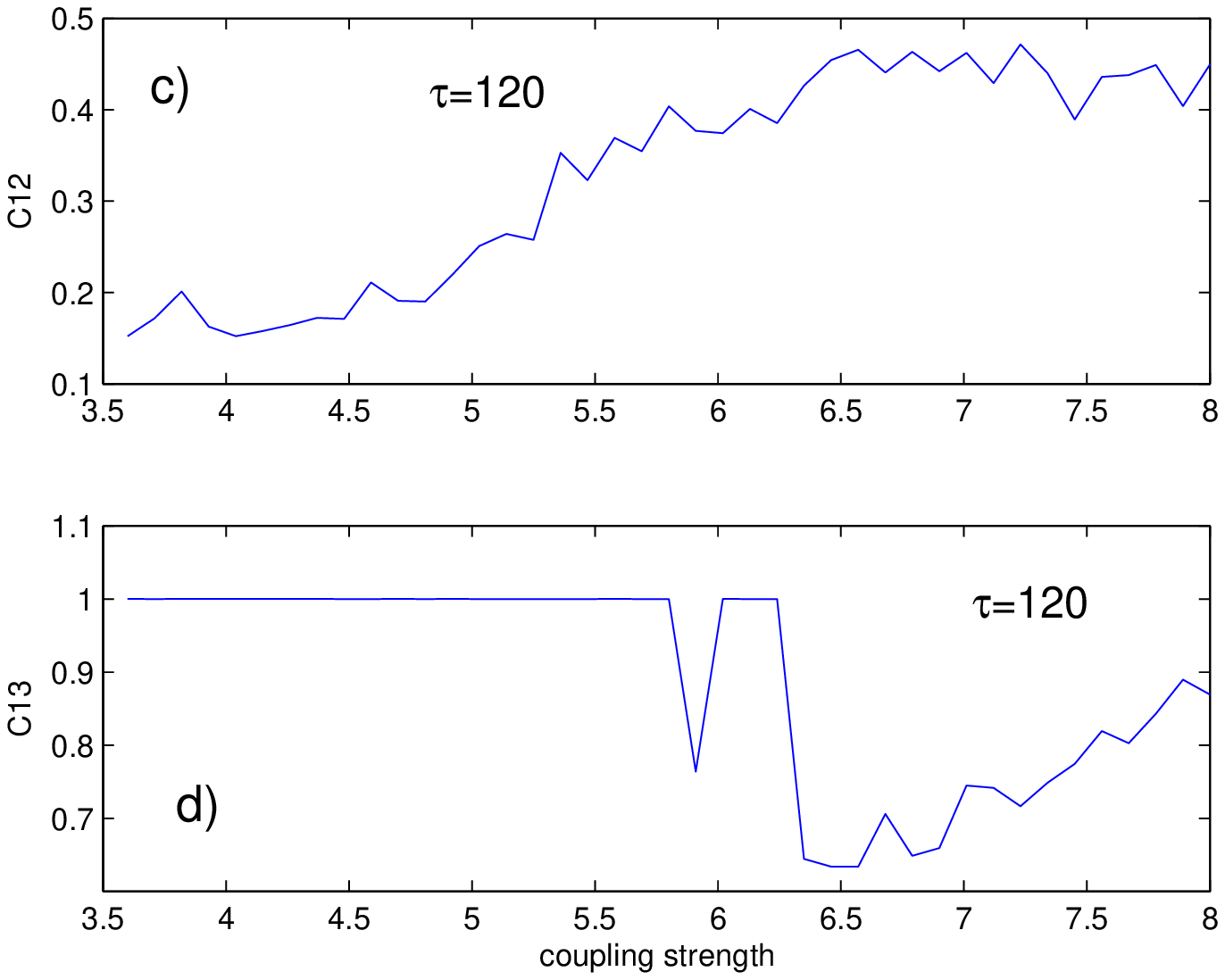}
}
\caption{Correlations corresponding to Fig. \ref{fig:SumLyapCoupling}.
a)  Correlation between the middle and one of the outer lasers, $\tau=60$.
b) Correlations of outer lasers, $\tau=60$.
c)  Correlation between the middle and one of the outer lasers, $\tau=120$.
d) Correlations of outer lasers, $\tau=120$.  
Outer lasers synchronize for greater range of
coupling strength as the delay is increased.  The middle and the outer
lasers show little correlation for all values of the coupling strengths.}
\label{fig:CorrCoupling}
\end{figure}

The amplitude of laser oscillations 
depends on the coupling strengths, $\delta_1$
and $\delta_2$, as well as the dissipation.  It was shown in the last
section that the product of the coupling strengths, $\delta_1 \delta_2$, has
to be strong enough to overcome the dissipation to cause the
onset of oscillations.   Increasing the coupling strengths increases
the role of nonlinearities in the system and the intensity of 
laser oscillations.  Since increased coupling pumps
more energy into the system, thereby increasing the effect of 
nonlinearities, the Lyapunov exponents may increase above zero, 
leading to desynchronization of the outer lasers.  In this
case, longer delays in coupling may be required in order for the outer lasers
to synchronize.  This effect is illustrated in Fig. \ref{fig:SumLyapCoupling},
which shows the sum of Lyapunov exponents as a function of coupling strengths
for two different delays, $\tau=60$ and $\tau=120$.  There is an 
abrupt increase in Lyapunov exponents above zero, due to increased
nonlinearity, as the coupling strength is increased. 
Increasing the delay however to $\tau=120$ leads to
synchronization for a greater range of coupling strengths, as compared
to $\tau=60$.  The corresponding correlations as a function of coupling
strengths are shown in Fig. \ref{fig:CorrCoupling}.  

It is worthwhile
to note that this loss of synchronization with increased coupling strengths
may seem counter-intuitive, and is not found in the case of fiber lasers
discussed in the following section.  Nevertheless,    
desynchronization at higher coupling strengths, and the synchronizing effect
of increased delays is in agreement with analytic results of this section.  
Since higher coupling strengths lead to greater fluctuations in
 $X(t)$, larger values of of $\tau$ or $\epsilon$ are needed in order
to satisfy Eqn. (\ref{eq:compare2}).  This means that increasing coupling
strength may lead to desynchronization unless the values of delay or
dissipation are increased accordingly.

The analysis in this section
focuses on the local stability of synchronized
state, rather than investigating the global properties of the system.
However, from numerical simulation, it appears that local stability
implies global stability, since the lasers synchronize, regardless of their
initial conditions, whenever the synchronous state is locally stable.
This suggests that the
system investigated in the present section only has a single attractor, unlike
the multiple attractor dynamics that can be induced in certain other systems
that have delayed feedback \cite{Zerega03}.  In
the presence of multiple attractors,
local stability of synchronized state would not necessarily result
in synchronization, since the initial condition can be such that the outer
systems end up in different attractor basins.  For synchronization in chaotic
systems with coexisting attractors, see \cite{Pisarchik06}.

\clearpage

\section{Synchronization in a Spatio-Temporal System: Fiber Lasers}

We next consider a different system with the same coupling
geometry of three delay-coupled components arranged linearly.  The
components are fiber ring lasers, which are a more complicated
system than semiconductor lasers.  A fiber ring laser contains of
a ring of optical fiber, a portion of which is doped and can lase.
Even a single, uncoupled ring laser is a time delayed system
because of the time that light takes to travel around the ring
(the round trip time), and a single ring laser can display
spatio-temporal chaos \cite{OjalvoR97}.  In contrast to the
semiconductor lasers, which have a much faster relaxation time,
the relaxation time scale of a fiber ring laser is on the order of
milliseconds to microseconds \cite{Lacot94,WilliamsGR97}.
Achieving a coupling delay that is long compared to the relaxation
time would require many kilometers of optical fiber in an
experiment and would be difficult to model because of
computational limitations. Instead, coupled fiber ring lasers are
often operated with a coupling delay that is short compared to the
relaxation time of the laser, and this is the case we address
here.  We include independent noise sources on each laser to
represent spontaneous emission.  The outer lasers are assumed to
be identical, while the center laser is detuned from the other
two.

We model the fiber lasers via a system of ordinary differential
equations for the population inversions coupled to a system of
maps for the electric field.  This model was used with other
coupling geometries in \cite{ShawSRR06,IBSLBSciteulike:1222889}
and is a variation of that introduced in \cite{WilliamsGR97}.

In each fiber ring laser, light circulates through a ring of
optical fiber, part of which is doped for stimulated emission. A
single polarization mode is modeled in each laser. Each laser is
characterized by a total population inversion $W(t)$ (averaged
over the length of the fiber amplifier) and an electric field
$E(t)$.  The time for light to circulate through the ring is the
cavity round trip time $\tau_R$.  Transit through the coupling
lines takes a potentially different delay time $\tau_d$.

The equations for the model dynamics are as follows:
\begin{eqnarray}
E_j(t) &=& R \exp \left[\Gamma (1-i\alpha_j) W_j(t)+i\Delta \phi\right] E^{\text{fdb}}_j(t) \nonumber \\
&& +\xi_j(t) \label{Eequ} \\
\frac{dW_j}{dt} &=& q -1-W_j(t) \nonumber \\
&& -\left|E^{\text{fdb}}_j(t)\right|^2 \left\{ \exp\left[2\Gamma W_j(t)\right]-1 \right\}, \label{Wequ}
\end{eqnarray}
The electric field from earlier times which affects the field at time $t$ is
\begin{eqnarray}
E^{\text{fdb}}_{1}(t)=E_{1}(t-\tau_R)+ \kappa E_{2}(t-\tau_d) \nonumber \\
E^{\text{fdb}}_{2}(t)=E_{2}(t-\tau_R)+\kappa E_{1}(t-\tau_d)+\kappa E_{3}(t-\tau_d) \label{Efdb} \\
E^{\text{fdb}}_{3}(t)=E_{3}(t-\tau_R)+\kappa E_{2}(t-\tau_d).
\nonumber
\end{eqnarray}
$E_j(t)$ is the complex envelope of the electric field in laser
$j$, measured at a given reference point inside the cavity.
$E^{\text{fbd}}_j(t)$ is a feedback term that includes optical
feedback within laser $j$ and optical coupling with the other
lasers.  Time is dimensionless, measured in units of the decay
time of the atomic transition, $\gamma_{||}^{-1}$.  The active
medium is characterized by the dimensionless detuning $\alpha_j$
between the transition and lasing frequencies and by the
dimensionless gain $\Gamma=\frac{1}{2}a L_a N_0$, where $a$ is the
material gain, $L_a$ the active fiber length, and $N_0$ the
population inversion at transparency.  The ring cavity is
characterized by its return coefficient $R$, which represents the
fraction of light remaining in the cavity after one round trip,
and the average phase change $\Delta \phi=2\pi n L_p/\lambda$ due
to propagation of light with wavelength $\lambda$ along the
passive fiber of length $L_p$ and index of refraction $n$.  Energy
input is given by the pump parameter $q$, which is measured in
units of the population decay rate $\gamma_{||}$.  The electric
field is perturbed by independent complex Gaussian noise sources
$\xi_j$ with standard deviation $D$.  Lasers 1 and 3 are each
coupled mutually with Laser 2 with a coupling strength of
$\kappa$, but Lasers 1 and 3 are not directly connected.  Values
of the parameters in the model, which are similar to those used in
an experiment for two coupled fiber lasers with self feedback
\cite{ShawSRR06}, are given in Table \ref{tab:parameters}.

\begin{table}
\caption{\label{tab:parameters}Parameters used in the coupled fiber laser model.}
\begin{ruledtabular}
\begin{tabular}{cccl}
Parameter & Value & Units & Description \\
\hline

$R$ & 0.4 & & output coupler return coefficient \\
$a$ & $2.03 \times 10^{-23}$ & m$^2$ & material gain coefficient \\
$L_a$ & 15 & m & length of active fiber \\
$L_p$ & 27 & m & length of passive fiber \\
$N_0$ & $10^{20}$ & m$^{-3}$ & transparency inversion \\
$\Gamma$ & 0.0152 & & dimensionless gain \\
$\alpha_1$ & 0.0202 & & detuning factor, laser 1 \\
$\alpha_2$ & 0.0352 & & detuning factor, laser 2 \\
$\alpha_3$ & 0.0202 & & detuning factor, laser 3 \\
$n$ & 1.44 & & index of refraction \\
$\lambda$ & $1.55\times 10^{-6}$ & m & wavelength \\
$\Delta \phi$ & $1.58\times 10^8$ & & average phase change \\
$D$ & 0.02 & & standard deviation of noise \\
$q$ & 100 & & pump parameter \\
$\gamma_{||}$ & 100 & s$^{-1}$ & population decay rate \\
$\tau_R$ & $201.6\times 10^{-9}$ & s & cavity round trip time \\
$\tau_d$ & $45\times 10^{-9}$ & s & delay time between lasers \\
$\kappa$ & 0-0.01 & & coupling strength
\end{tabular}
\end{ruledtabular}
\end{table}

Eqns.~\ref{Eequ}-\ref{Wequ} consist of a delay differential equation for 
$W(t)$ coupled to a map for $E(t)$.  We integrated Eqn.~\ref{Wequ} 
numerically using Heun's method while propagating the map in 
Eqn.~\ref{Eequ}.  The time step for integration was 
$\tau_R/N$, where $N=600$.  This step size corresponds to dividing the ring cavity into $N$ spatial elements.

Because of the feedback term $E^{\text{fdb}}_j(t)$ in Eqn.~\ref{Eequ}, one 
can think of Eqn.~\ref{Eequ} as mapping the electric field on the time interval $[t-\tau_R,t]$ to the time interval $[t,t+\tau_R]$ in the absence of coupling ($\kappa=0$).  Equivalently, because the light is traveling around the cavity, Eqn.~\ref{Eequ} maps the electric field at all points in the ring at time $t$ to the electric field at all points in the ring at time $t+\tau_R$.  We can thus construct spatio-temporal plots for $E(t)$ or the intensity $I(t)=\left| E(t) \right|^2$ by unwrapping $E(t)$ into segments of length $\tau_R$.

To correspond with previous experiments in which the measured
light intensity is passed through a 125 MHz bandwidth
photodetector \cite{ShawSRR06}, we computed intensities from model
and applied a low pass filter with $f_0=125$ MHz, multiplying the
Fourier transform by the transfer function
\begin{equation}
G=\left\{ \left(i \frac{f}{f_0}+1\right) \left[-\left(\frac{f}{f_0}\right)^2+i \frac{f}{f_0}+1\right] \right\} ^{-1}.
\end{equation}
All results presented here are based on the filtered intensity.

\begin{figure}
\includegraphics[
 width=5in,
 keepaspectratio]{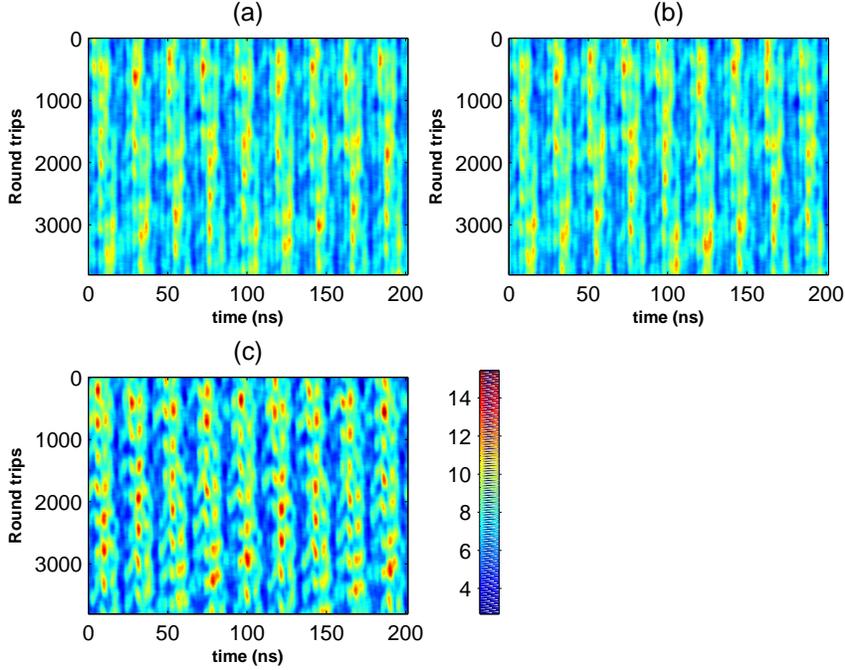}
\caption{Spatio-temporal plots of intensity for three lasers coupled in a 
line with $\kappa=0.005$.  (a) Laser 1, (b) Laser 3, (c) Laser 2.  All plots have the same color scale.}
\label{fig:stplots}
\end{figure}

The coupled fiber laser model can exhibit several types of
dynamics.  Long time scale behavior is most easily seen through
spatio-temporal plots, in which the time series is unwrapped into
intervals of one round trip $\tau_R$, and subsequent round trips
are stacked on top of each other.  An example is shown in
Fig.~\ref{fig:stplots} for a coupling strength of $\kappa=0.005$.
The dynamics approximately repeats from one round trip to the next
but evolves gradually over tens or hundreds of round trips.  At
this coupling value, similarities between the outer lasers (1
and 3) can be seen.  Other coupling strengths can produce behavior
that appears to be noisy periodic.

\begin{figure}
\includegraphics[width=5in,keepaspectratio]{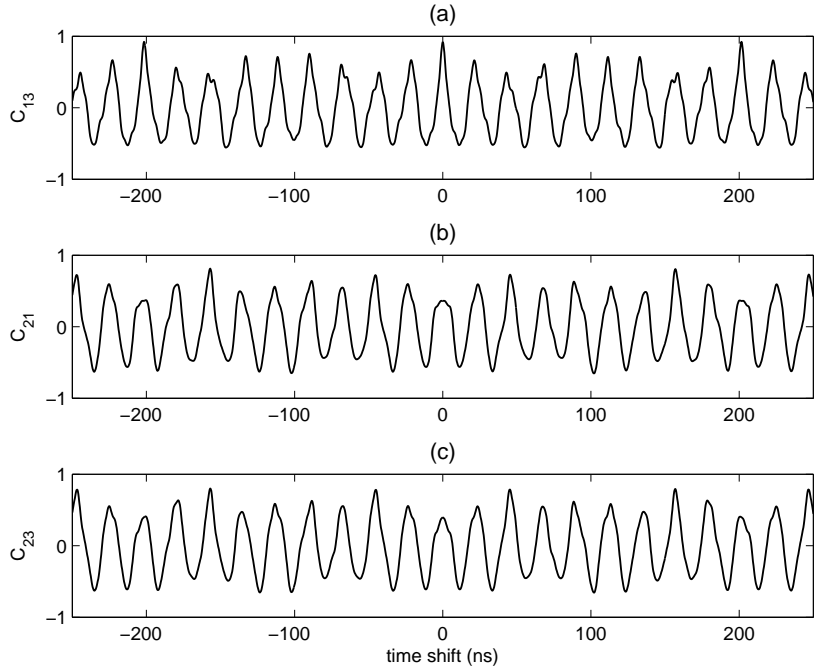}
\caption{Cross correlation between lasers vs.~time shift between the 
laser 
time series.  $\kappa=0.005$, and averaging was done over 10 round trips.  Time series were taken near the beginning of Fig.~\ref{fig:stplots}. (a) Laser 1 and Laser 3, (b) Laser 2 and Laser 1, (c) Laser 2 and Laser 3.}
\label{fig:cc_vs_shift}
\end{figure}

To assess the type and extent of synchronization in the system, we
shift the laser time series relative to each other and compute
cross correlations between them.  An example is given in
Fig.~\ref{fig:cc_vs_shift}.  Lasers 1 and 3, the outer lasers,
have a peak in the cross correlation at a time shift of zero,
meaning they are isochronally synchronized.  Other maxima occur at
multiples of the round trip time $\tau_R$, 201.6 ns, because the
laser time series approximately repeat every round trip.  Laser 2
is not isochronally synchronized with the outer lasers; the peak
at zero time shift is small.  However, there are more significant
peaks in the correlation between Laser 2 and the outer lasers at a
shift of the coupling time $\tau_d$, i.e., $\pm 45$ ns, indicating
partial delay synchrony.  The cross correlation is approximately
equal whether the lasers are compared with Laser 2 leading the
others or with Laser 2 following them.  This result is consistent
with the reversible delay synchrony observed previously
for two mutually delay coupled fiber ring lasers, for which there
is no clear leader and follower \cite{ShawSRR06}.

\begin{figure}
\includegraphics[
 width=3in,
 keepaspectratio]{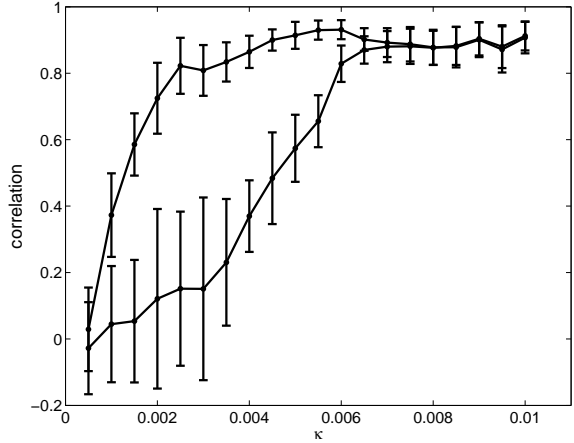}
\caption{Cross correlation vs.~coupling for three lasers coupled in a 
line.  Upper curve:  Laser 1 and Laser 3 compared with zero time shift.  Lower curve:  Laser 1 compared to Laser 2 with a lag of the coupling delay time.  Relationship between Lasers 3 and 2 is similar to that between 1 and 2.}
\label{fig:cc_vs_coupling}
\end{figure}

We next determine the long time synchronization behavior of the
lasers.  For each round trip, we compute the cross correlation
between outer lasers without a time shift and between the center
and outer lasers with a time shift of $\tau_d$.  We shift so
that Laser 2 leads the others, but because of the reversible
synchronization, we would obtain similar results for cross
correlations with Laser 2 following the others.  To obtain good
statistics, we average the round trip cross correlations over five
separate intervals of 8 ms each, which are spaced apart by 100 ms.
The standard deviation over all the round trips serves as an error
estimate. Fig.~\ref{fig:cc_vs_coupling} shows how the
synchronization depends on the coupling strength.  The outer
lasers begin to synchronize isochronally at weak coupling and are
well synchronized by the time $\kappa$ reaches 0.5\%.  Delay
synchrony between center and outer lasers requires a stronger
coupling, but the level of delay synchrony eventually saturates to
the same level as that of the isochronal synchrony between outer
lasers. It is likely that the delay synchrony arises more slowly
because Laser 2 is detuned away from the others, while Lasers 1
and 3 are identical.  Perfect synchrony is not achieved due to the
noise in the system.

To relate the fiber laser system to the coupled semiconductor
lasers discussed in the previous section, one might consider
whether increasing coupling delay or increasing dissipation will
improve synchrony in the fiber lasers.  In the regime where the
coupling delay is on the same order as the round trip time, and
much less than the laser relaxation time, the delay has little
effect on the synchronization.  Increasing dissipation by
increasing the decay rate  $\gamma_{||}$ does not improve
synchronization either.  The fiber lasers behave differently than
the semiconductor lasers in several respects, although the
coupling geometry with three lasers in a line leads to synchrony
of the outer lasers in both cases.

\begin{figure}
\includegraphics[
 width=3in,
 keepaspectratio]{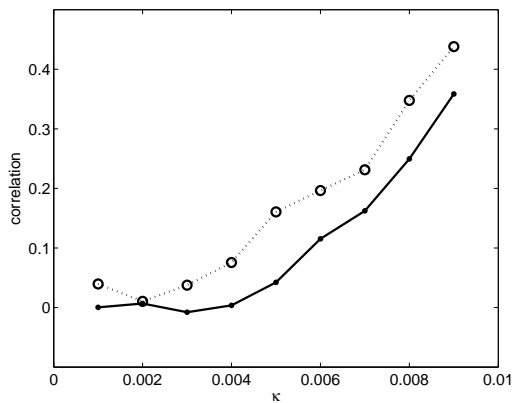}
\caption{Cross correlation vs.~coupling for unidirectionally driven lasers 
(Laser 2 driving Lasers 1 and 3).  Solid line and filled points:  Laser 1 and Laser 3 compared with zero time shift.  Dashed line and open circles:  Laser 1 compared to Laser 2 with a lag of the coupling delay time.}
\label{fig:cc_unidir}
\end{figure}

Since fiber lasers had short coupling delays, with respect to 
relaxation frequency, mutual coupling might be expected to play an
important role in synchronization.  
To compare the mutually coupled and the purely
driven geometries, we simulated the case of 
Lasers 1 and 3 being driven by a common input, Laser 2, but with no 
feedback from Lasers 1 and 3 into Laser 2.  Synchronization results are 
given in Fig.~\ref{fig:cc_unidir}.  These results were computed in the 
same way as for Fig.~\ref{fig:cc_vs_coupling}.  Isochronal synchrony does 
begin to occur between Lasers 1 and 3 due to their common input, but they 
require a stronger coupling than in the previous case.  It appears that 
generalized synchrony plays a role in the synchronization of the outer 
fiber lasers, but the mutual coupling between the outer and center lasers 
is also important.

\section{Discussion}

Using three mutually delay coupled lasers in a line, we have shown that
synchronization exists in two different laser systems as a result of
coupling architecture. The two types of lasers we have considered are
semiconductor lasers and fiber ring lasers. Individually, the
semiconductor lasers are low dimensional (as represented by intensity and
population inversion), while the fiber ring lasers are considered
spatio-temporal since there exist on the order of 1000 modes coupled
within each laser.

Since the coupling is done using finite lengths of fiber, the lasers
communicate with finite delay.  Due to symmetric coupling, a solution
exists where the dynamics of the outer lasers are identical.  If this
symmetric solution has global stability, then the outer lasers will become
synchronized with zero lag.  This architecture is in contrast to two 
lasers mutually coupled with delay, for which the solutions are 
synchronized but with a lag equal to the coupling delay.

In general, stability of the zero lag synchronized state is difficult to
show for the case of nonlinear mutually coupled systems.  However, we have
been able to analyze the local stability of the synchronized state for the
case of coupled semiconductor lasers with long delays, where the delay is
typically long compared to the relaxation oscillation frequency.   
In this case, it can be demonstrated that
synchronization is explicitly dependent on dissipation in the internal
dynamics of the outer lasers, so that synchronization is due to ``washing
out'' of the difference in initial conditions.

It is worthwhile to note that this sort of synchronization where
the sum of Lyapunov exponents has a negative linear dependence on dissipation
is also seen in the context of generalized synchronization in driven
dissipative systems.  In this case, there is unidirectional coupling from the
driver to the response system with the onset of generalized synchronization
whenever the dynamics of the response system become a function of the
driver.  While in the case of mutually coupled semiconductor lasers with
long delays, it is clear that the signal the outer lasers recieve from
the center laser is affected by mutual coupling, the dependence of
transverse Lyapunov stability 
on dissipation in the outer lasers indicates
that the outer lasers synchronize due to a common signal from the
middle laser.  This explains the linear dependence of transverse
Lyapunov exponents on dissipation, since two identical response systems
will synchronize to the common signal from the driver, provided there
is some internal dissipation in the response systems themselves to 
``wash out'' any difference in initial conditions. 
It is perhaps not too surprising that for sufficiently long delays the chaotic 
synchronization phenomena may appear to
be similar to driven systems, since on the time scale of the delay time, the
dynamics of the outer lasers do not affect the mutual coupling term in their
equations, just as the driven system cannot affect the signal it receives
from the driver.  It follows that for sufficiently long delays, the middle
laser can be viewed as driving the dynamics of the outer lasers.  Then,
complete synchronization of the outer lasers is the result of generalized
synchronization between the middle and the outer lasers.  The cases where
the outer lasers seem to anticipate the dynamics of the middle laser (due
to the detuning values used), need not be excluded since anticipatory
synchronization is seen in purely driven systems as well, as was discussed
in the introduction to this chapter. 

While for coupled semiconductor systems, chaotic synchronization seems in
general to improve with longer delays, long delays are not necessary to
obtain robust synchronization.  There are many examples of synchronization
in mutually coupled systems without delays, including semiconductor lasers.  
Fiber lasers serve as another
example of a laser system that synchronizes for relatively short delays,
where the coupling delay is short compared to the relaxation oscillation
frequency.  In contrast to semiconductor lasers, which will settle into a
steady state in the absence of mutual coupling, fiber lasers will continue
to oscillate, sometimes showing complicated dynamics even when left
uncoupled. This is due to intrinsic noise from spontaneous emission in
each laser.  Further research is needed to elucidate the
phenomena behind fiber laser synchronization.  One possible approach may
be to model phase synchronization in the limit of weak coupling, in the
regime where mutual coupling affects only the phase and not the amplitude
of the lasers. 

There are also a number of questions that remain to be answered in regard
to synchronization in semiconductor lasers.  One interesting question is
the effect of delays on synchronization.  While we have shown that for
sufficiently long delays, optoelectronically coupled lasers will
synchronize, it remains to be explained why shortening the delays
sometimes leads to desynchronization.  This issue becomes especially
interesting when the delays are still long compared to the oscillation
time but not long enough to result in synchronization.  In this case,
some non-trivial resonance-like phenomenon may be occurring, which requires
further investigation
A related question that could be addressed is the decorrelation
time of the signals as the coupling delay is increased.  Thus for short
coupling delays in semiconductor lasers, the dynamics are more regular,
and synchronization may indeed hinge on the regularity of
the signal and self-correlations in the
dynamics.  In this case, the generalized synchronization idea may not
play an important role, since the ``driving signal'' from the middle laser
adjusts itself quickly to the current dynamics of the outer lasers.  
At longer delays, however, the dynamics become decorrelated,
and some measure of mutual information would be beneficial to quantify
this phenomena.  

Further study of both types of laser systems discussed here may lead to a
better understanding of the key requirements for synchronization and the
important role played by the linear coupling architecture.  In addition,
these ideas may help in designing coupling architectures to synchronize 
larger numbers of lasers.

We gratefully acknowledge support from the Office of Naval Research.  LBS 
and ALS are currently National Research Council postdoctoral fellows.

\end{document}